\documentclass[journal]{vgtc}                % final (journal style)
\ifpdf%                                % if we use pdflatex
  \pdfoutput=1\relax                   % create PDFs from pdfLaTeX
  \usepackage{graphicx}                % allow us to embed graphics files
  \DeclareGraphicsExtensions{.pdf,.png,.jpg,.jpeg} % for pdflatex we expect .pdf, .png, or .jpg files
\else%                                 % else we use pure latex
  \usepackage{graphicx}                % allow us to embed graphics files
  \DeclareGraphicsExtensions{.eps}     % for pure latex we expect eps files
\fi%

\graphicspath{{figures/}{pictures/}{images/}{./}} % where to search for the images

\usepackage{microtype}                 % use micro-typography (slightly more compact, better to read)
\PassOptionsToPackage{warn}{textcomp}  % to address font issues with \textrightarrow
\usepackage{textcomp}                  % use better special symbols
\usepackage{mathptmx}                  % use matching math font
\usepackage{times}                     % we use Times as the main font
\usepackage{cite}                      % needed to automatically sort the references
\usepackage{tabu}                      % only used for the table example
\usepackage{booktabs}                  % only used for the table example
\usepackage{balance}

\onlineid{0}

\vgtccategory{Research}
\vgtcpapertype{please specify}

\usepackage{color}
\usepackage{xcolor}
\usepackage{epsfig}
\usepackage{graphicx}
\usepackage{amsmath}
\usepackage{amssymb}

\usepackage{textcomp}
\usepackage{makecell}
\usepackage{subcaption}
\usepackage{tabularx}
\usepackage{array}
\newcolumntype{L}[1]{>{\raggedright\let\newline\\\arraybackslash\hspace{0pt}}m{#1}}
\newcolumntype{C}[1]{>{\centering\let\newline\\\arraybackslash\hspace{0pt}}m{#1}}
\newcolumntype{R}[1]{>{\raggedleft\let\newline\\\arraybackslash\hspace{0pt}}m{#1}}
\usepackage{lipsum}
\newcommand\blfootnote[1]{%
  \begingroup
  \renewcommand\thefootnote{}\footnote{#1}%
  \addtocounter{footnote}{-1}%
  \endgroup
}

\newcommand{\tabfig}[2]{\parbox[c]{1.5em}{\includegraphics[width=#1 in]{#2}}}

\title{Scene-Aware Audio Rendering via Deep Acoustic Analysis}
\author{Zhenyu Tang, Nicholas J. Bryan, Dingzeyu Li, Timothy R. Langlois, and Dinesh Manocha}
\authorfooter{
\item
 Zhenyu Tang and Dinesh Manocha are with the University of Maryland. E-mail: \{zhy,dm\}@cs.umd.edu.
\item
 Nicholas J. Bryan, Dingzeyu Li and Timothy R. Langlois are with Adobe Research. E-mail: \{nibryan,dinli,tlangloi\}@adobe.com.
}

\shortauthortitle{Tang \MakeLowercase{\textit{et al.}}: Scene-Aware Audio Rendering via Deep Acoustic Analysis}
\abstract{
We present a new method to capture the acoustic characteristics of real-world rooms using commodity devices, and use the captured characteristics to generate similar sounding sources with virtual models. Given the captured audio and an approximate geometric model of a real-world room, we present a novel learning-based method to estimate its acoustic material properties.  Our approach is based on deep neural networks that estimate the reverberation time and equalization of the room from recorded audio. These estimates are  used to compute material properties related to room reverberation using a novel material optimization objective. We use the estimated acoustic material characteristics for audio rendering using  interactive geometric sound propagation and highlight the performance on many real-world scenarios. We also perform a user study to evaluate the perceptual similarity between the recorded sounds and our rendered audio.
} % end of abstract

\keywords{Audio rendering, audio learning, material optimization.}

\CCScatlist{ % not used in journal version
 \CCScat{K.6.1}{Management of Computing and Information Systems}%
{Project and People Management}{Life Cycle};
 \CCScat{K.7.m}{The Computing Profession}{Miscellaneous}{Ethics}
}

\teaser{
  \centering
\includegraphics[width=\linewidth]{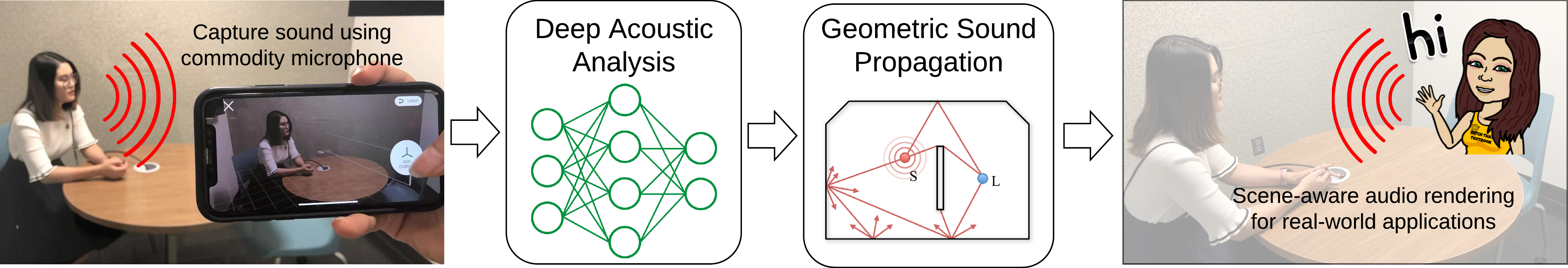}
\caption{Given a natural sound in a real-world  room  that is recorded using a cellphone microphone (left), %and the rough geometry that can be captured by smartphone cameras, 
we estimate the acoustic material properties and the frequency equalization of the room using a novel deep learning approach (middle). We use the estimated acoustic material properties for generating plausible sound effects  in the virtual model of the room (right). Our approach is general and robust, and works well with commodity devices.
}
\label{fig:teaser}
}

\begin{document}


\firstsection{Introduction}

\maketitle
Auditory perception of recorded sound is strongly affected by the acoustic environment it is captured in. 
Concert halls are carefully designed to enhance the sound on stage, even accounting for the effects an audience of human bodies will have on the propagation of sound~\cite{barron_2010}. 
Anechoic chambers are designed to remove acoustic reflections and propagation effects as much as possible. 
Home theaters are designed with acoustic absorption and diffusion panels, as well as with careful speaker and seating arrangements~\cite{rizzi2016small}. 

The same acoustic effects are important when creating immersive effects for virtual reality (VR) and augmented reality (AR) applications. 
It is well known that realistic sounds can improve a user's sense of presence and immersion~\cite{larsson2002}. 
There is considerable work on interactive sound propagation in virtual environments based on geometric and wave-based methods~\cite{vorlander1989,schissler2017interactive,raghuvanshi2014parametric,cao2017bidirectional}. 
Furthermore, these techniques are increasingly used to generate plausible sound effects in VR systems and games, including
%Video game makers simulate audio to increase immersion and react to user actions, through efforts such as 
Microsoft Project Acoustics\footnote{https://aka.ms/acoustics}, Oculus Spatializer\footnote{https://developer.oculus.com/downloads/package/oculus-spatializer-unity}, and Steam Audio\footnote{https://valvesoftware.github.io/steam-audio}. However, these methods are limited to synthetic scenes where an exact geometric representation of the scene and acoustic material properties are known apriori.
%and audio simulation modules in Unity and Unreal Engine. 

In this paper, we address the problem of rendering realistic sounds that are similar to recordings of real acoustic scenes. \blfootnote{Project website \url{https://gamma.umd.edu/pro/sound/sceneaware}}
These capabilities are needed for VR as well as AR applications~\cite{AES2018}, which often use recorded sounds.
%including rendering virtual, animated avatars into real-world audio-visual scenes for futuristic telecommunication systems and/or content creation. .
Foley artists often record source audio in environments similar to the places the visual contents were recorded in. 
%Video game makers simulate audio to increase immersion and react to user actions, through efforts such as Project Acoustics\footnote{https://aka.ms/acoustics}, Oculus Spatializer\footnote{https://developer.oculus.com/downloads/package/oculus-spatializer-unity}, Steam Audio\footnote{https://valvesoftware.github.io/steam-audio} and audio simulation modules in Unity and Unreal Engine. 
Similarly, creators of vocal content (e.g. podcasts, movie dialogue, or video voice-overs), carefully re-record content made in different environment or with different equipment to match the acoustic conditions. 
However, these processes are expensive, time-consuming, and cannot adapt to spatial listening location.
There is strong interest in developing automatic spatial audio synthesis methods.

For VR or AR content creation, acoustic effects can also be captured with an impulse response (IR) -- a compact acoustic description of how sound propagates from one location to another in a given scene. A given IR can be convolved with any virtual sound or dry sound to generate the desired acoustic effects.
%To apply an IR and simulate sound propagation between two points, one convolves an input sound source with the IR. 
%IR-based convolution reverberation, however, becomes difficult when a desired IR is unavailable or is inconvenient to obtain. 
%In AR applications, for example, the acoustic environments are not known before hand, so %pre-computation of the IR is impractical.
However, recording the IRs of real-world scenes can be challenging, especially for interactive applications. Many times special recording hardware is needed to record the IRs. 
Furthermore, the IR is a function of the source and listener positions and it needs to be re-recorded as either position changes.
%on-the-fly, however, is impractical. Requiring manual IR measurements  destroys the real-time aspect necessary for interactive systems. IR measurements also typically require quiet acoustic environments, which is disruptive to other people in the environment. 
%It is sometimes difficult to gain access to the environment the video was recorded in (perhaps it was a set that was dismantled, or a remote location). 
%Additionally, an IR is a function of position, so it would need to continuously change as the source and/or listener move.

Our goal is to replace the step of recording an IR with an unobtrusive method that works on in-situ speech recordings and video signals and uses commodity devices.
This can be regarded as an acoustic analogy of visual relighting~\cite{debevec2002image}: 
to light a new visual object in an image, traditional image based lighting methods require the capture of real-world illumination as an omnidirectional, high dynamic range (HDR) image. 
This light can be applied to the scene, as well as on a newly inserted object, making the object appear as if it was always in the scene.
Recently, Gardner et al.~\cite{gardner2017learning} and Hold-Geoffroy et al.~\cite{hold2017deep} proposed convolutional neural network (CNN)-based methods to estimate HDR indoor or outdoor illumination from a single low dynamic range (LDR) image.
These high-quality visual illumination estimation methods enable novel interactive applications.
Concurrent work from LeGendre et al.~\cite{legendre2019deeplight} demonstrates the effectiveness on mobile devices, enabling photorealistic mobile mixed reality experiences.

\begin{figure*}[t]
\centering 
\includegraphics[width=0.8\linewidth]{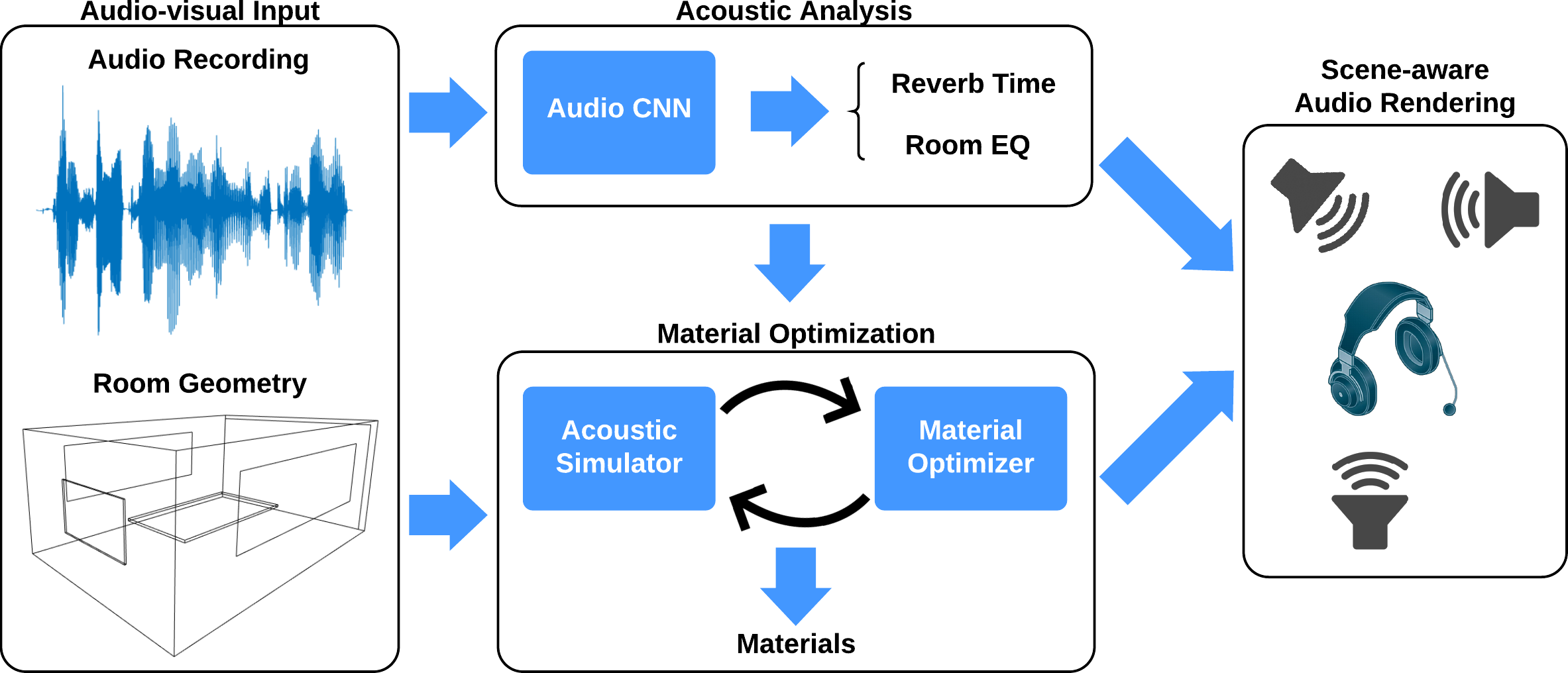}
\caption{{\bf Our pipeline:} Starting with an audio-video recording (left), %which can be captured by a smartphone,
we estimate the 3D geometric representation of the environment using standard computer vision methods. 
We use the reconstructed 3D model to simulate new audio effects in that scene.
To ensure our simulation results perceptually match recorded audio in the scene, we automatically estimate two acoustic properties from the audio recordings: frequency-dependent reverberation time or $T_{60}$ of the environment, and a frequency-dependent equalization curve. 
The $T_{60}$ is used to optimize the frequency-dependent absorption coefficients of the materials in the scene. 
The frequency equalization filter is applied to the simulated audio, and accounts for the missing wave effects in geometrical acoustics simulation. We use these parameters for interactive scene-aware audio rendering (right).
\label{fig:pipeline}
}
\end{figure*}

% \todo{coming back to audio relighting, show that there are some work along this line, but they are all limited}
In terms of audio ``relighting" or reproduction, 
%\zhy{I changed  to ``reproduction" as this is an established term}, 
there have been several approaches proposed toward 
realistic audio in 360\textdegree{} images~\cite{kim2019immersive}, 
multi-modal estimation and optimization~\cite{schissler2017acoustic}, 
and scene-aware audio in 360\textdegree{} videos~\cite{Li:2018:360audio}.
However, these approaches either require separate recording of an IR, or produce audio results that are perceptually different from recorded scene audio.
%IRs have been the main way to capture and compute acoustic properties. 
Important acoustic properties can be extracted from IRs, including the reverberation time ($T_{60}$), which is defined as the time it takes for a sound to decay 60 decibels~\cite{kuttruff2016room}, and the frequency-dependent amplitude level or equalization (EQ)~\cite{hak2012measuring}.
%This heavy reliance on IRs greatly constrains the wide adoption of audio for immersive applications or video post-production that require realistic acoustic simulation that is calibrated to real-world acoustic scenes.

\vspace*{0.1in}
% \zhy{Can we move paragraphs below until summary/contribution to after presenting our main contributions? We've put too many technical contents in the intro section without defining a lot of terms.}
% \todo{we solve this problem without requiring the IR}
\noindent {\bf Main Results:}
We present novel algorithms to estimate two important environmental acoustic properties from recorded sounds (e.g. speech). 
Our approach uses commodity microphones and does not need to capture any IRs. 
The first property is the frequency-dependent $T_{60}$. 
This is used to optimize absorption coefficients for  geometric acoustic (GA) simulators for audio rendering. 
Next, we estimate a frequency equalization filter to account for wave effects that cannot be modeled accurately using geometric acoustic simulation algorithms. 
This equalization step is crucial to ensuring that our GA simulator outputs perceptually match existing recorded audio in the scene.

%\todo{We show that XXX (something simliar to \cite{legendre2019deeplight}.}
%\todo{Figure~\ref{fig:pipeline}}

%This allows us to simulate audio in an environment (accounting for source/listener movement), and also \todo{blend} well with recorded audio.

%\todo{another paragraph on why it's hard}
Estimating the equalization filter \emph{without an IR} is challenging since it is not only speaker dependent, but also scene dependent, which poses extra difficulties in terms of dataset collection.
%Blind estimation of acoustic properties and parameters is an emerging research area;
%however, all existing efforts have focused on $T_{60}$ or other acoustic properties, where the early and late parts of the speech can be used together. % \todo{confusing sentence}.
%In addition to $T_{60}$ , we estimatEqualization, on the other hand, is not a time-dependent phenomenon. 
For a model to predict the equalization filtering behavior accurately, we need a large amount of diverse speech data and IRs. 
Our key idea is a novel dataset augmentation process that significantly increases room equalization variation. 
% \todo{do we really increase speaker variety? or just equalizationin variety?} \zhy{Might be an overstatement about speaker variety. We did intentionally split datasets so each set sees different speakers. But I think that's the basic.}
%\todo{Talk about the new optimization goal}
With robust room acoustic estimation as input, we present a novel inverse material optimization algorithm to estimate the acoustic properties.
%further improve the state-of-the-art inverse material optimization algorithm~\cite{Li:2018:360audio}.
%Typical off-the-shelf simulators can produce ``plausible'' audio when presented in isolation with a virtual or real scene.  In our applications we seek to recreate or match the same room acoustics of existing acoustical spaces. This requires us to provide the right simulator parameters to produce ``authentic'' audio that is calibrated to real-world sound scenes~\cite{kim2019immersive}.
We propose a new objective function for material optimization and show that it models the IR decay behavior better than the technique by Li et al.~\cite{Li:2018:360audio}.
We demonstrate our ability to add new sound sources in regular videos. 
Similar to visual relighting examples where new objects can be rendered with photorealistic lighting, we enable audio reproduction in any regular video with existing sound with applications for mixed reality experiences. We highlight their performance on many challenging benchmarks.

We show the importance of matched $T_{60}$ and equalization in our perceptual user study \S\ref{sec:study}. In particular, our perceptual evaluation results show that: (1) Our $T_{60}$ estimation method is perceptually comparable to all past baseline approaches, even though we do not require an explicit measured IR; (2) Our EQ estimation method improves the performance of our $T_{60}$-only approach by a statistically significant amount ($\approx$ 10 rating points on a 100 point scale); and (3) Our combined method ($T_{60}$+EQ) outperforms the average room IR ($T_{60}=.5$ seconds with uniform EQ) by a statistically significant amount ($+10$ rating points) -- the only reasonable comparable baseline we could conceive that does not require an explicit IR estimate.
To the best of our knowledge, ours is the first method to predict IR equalization from raw speech data and validate its accuracy.
Our main contributions include:
% \todo{write a short paragraph on user study}
% We conducted a user study to validate 
% our acoustic property estimator and the optimized renderer.
% %In \S\ref{sec:app}, we show the the seamless blending of new audio to existing audio and validate the quality with a user study.
% %We also showcase exciting new application in virtual avatar in video production and remote conferencing.
% We carefully designed and carried out two separate critical listening tests, one on authenticity and the other on plausibility.
% In each test, we studied the perceptual similarity of a reference speech recording compared to speech recordings convolved with simulated impulse responses.  
%In summary, we highlight the following contributions:
\begin{itemize}
  \setlength\itemsep{0em}
    \item A CNN-based model to estimate frequency-dependent $T_{60}$ and equalization filter from real-world speech recordings.
    % \item A CNN-based model to estimate a frequency-dependent equalization filter from a real-world speech recordings.
    % I combined them again since it would seem too many duplicate text otherwise.
    \item An equalization augmentation scheme for training to improve the prediction robustness.
    \item A derivation for a new optimization objective that better models the IR decay process for inverse materials optimization.
    \item A user study to compare and validate our performance with current state-of-the-art audio rendering algorithms. Our study is used to evaluate the perceptual similarity between the recorded sounds and our rendered audio.
\end{itemize}
%To present our contributions, we discuss related work, outline our method, discuss analysis and applications, justify our work with perceptual evaluation, and finally conclude.

%\todo{summarize contributions}
% \zhy{Let's rewrite them into bullet point contributions (mention user study results too)?}
% In summary, we introduce a CNN-based method to predict sub-band $T_{60}$ and equalization from a real-world speech recording.
% To address the dataset collection challenge, we strategically design an equalization augmentation scheme.
% We also propose a new optimization objective that better models the IR decay process.
% With our complete pipeline (Figure~\ref{fig:pipeline}), we can now author immersive audio on regular video inputs and even mixed reality applications.
% \paragraph{Main contributions}
% \begin{enumerate}
% 	\item One
% 	\item Two
% 	\item Three
% \end{enumerate}

%\todo{from Zhenyu: contribution should be coming from using the correct dataset to having trained the models as a whole, not just magnifying the  network design.}

% Our contributions are 
% \begin{itemize}
%     \item Network to predict modulation
%     \item Modulation augmentation
%     \item anything on fitting with AR video
% \end{itemize}

%%%%%%%%%%%%%%%%%%%%%%%%%%%%%%%%%%%%%%%%%%%%%%%%%%%%%%%%%%%%%%%%%%%%%%%%%%%%%%%%
%%%%%%%%%%%%%%%%%%%%%%%%%%%%%%%%%%%%%%%%%%%%%%%%%%%%%%%%%%%%%%%%%%%%%%%%%%%%%%%%
%%%%%%%%%%%%%%%%%%%%%%%%%%%%%%%%%%%%%%%%%%%%%%%%%%%%%%%%%%%%%%%%%%%%%%%%%%%%%%%%
\section{Related Work}

% immersive audio / scene-aware audio
% \todo{put plausibility/authenticity in intro?}
%\zhy{cite some psycho-acoustic papers since this topic relates to perception?} 
% I think this is discussed in Kim, which is cited below.
% we don't have to find a reference for every sentence or statement.
Cohesive audio in mixed reality environments (when there is a mix of real and virtual content), is more difficult than in fully virtual environments.
This stems from the difference between ``Plausibility'' in VR and ``Authenticity'' in AR~\cite{kim2019immersive}. 
Visual cues dominate acoustic cues, so the perceptual difference between how audio sounds and the environment in which it is seen is smaller than the perceived environment of two sounds. %\zhy{need references?} \todo{this is described in ref 24, Kim et al. 2019}.
Recently, Li et al.\ introduced scene-aware audio to optimize simulator parameters to match the room acoustics from existing recordings~\cite{Li:2018:360audio}.
By leveraging visual information for acoustic material classification, Schissler et al.\ demonstrated realistic audio for 3D-reconstructed real-world scenes~\cite{schissler2017acoustic}.
However, both of these methods still require explicit measurement of IRs. 
In contrast, our proposed pipeline works with any input speech signal and commodity microphones.

%\paragraph{Sound Sim}
Sound simulation can be categorized into wave-based methods and geometric acoustics. 
While wave-based methods generally produce more accurate results, it remains an open challenge to build a real-time universal wave solver.
Recent advances such as parallelization via rectangular decomposition~\cite{morales2015parallel},
pre-computation acceleration structures~\cite{mehra2015wave},
and coupling with geometric acoustics~\cite{yeh2013wave,rungta2018diffraction} are used for interactive applications.
It is also possible to precompute low-frequency wave-based propagation effects in large scenes~\cite{Raghuvanshi:2010:PWS}, and to perceptually compress them to reduce 
runtime requirements~\cite{Raghuvanshi:2014:PWF}. Even with the massive speedups presented, and a real-time runtime engine, these methods still require
tens of minutes to hours of pre-computation depending on the size of the scene and frequency range chosen, making them impractical for
augmented reality scenarios and difficult to include in an optimization loop to estimate material parameters.
With interactive applications as our goal, most game engines and VR systems tend to use geometric acoustic simulation methods~\cite{vorlander1989,schissler2017interactive,cao2017bidirectional,schissler2018interactive}. These algorithms are based on fast ray tracing and perform specular and diffuse reflections~\cite{savioja2015overviewGA}. Some techniques have been proposed to approximate low-frequency diffraction effects using ray-tracing~\cite{tsingos2001,rungta2018diffraction,Taylor12}. Our approach can be combined with any interactive audio simulation method, though our current implementation is based on bidirectional ray tracing~\cite{cao2017bidirectional}. The sound propagation algorithms can also be used for acoustic material design optimization for synthetic scenes~\cite{morales2016efficient}. 

%based on interactive geometric propogation algorithms, though 
%For an overview of geometrical acoustic methods, please see the work of Savioja and Svensson~\cite{savioja2015overviewGA}.
%Our proposed method can be incorporated into any GA-based simulator, such as~\cite{schissler2017interactive,rungta2018diffraction}.

% \todo{maybe cite a few more GA papers and say that our method is not tied to this specific one?}

% \todo{write about room equliazation}
% Room resonances from wave effects like standing waves has been a long standing challenge~\cite{groh1974high}.
% Karjalainen et al. discussed the effects of resonance in room equalization and dereverberation~\cite{karjalainen2005room}. 
% \cite{corbach2009automated}

%\paragraph{CNN Audio Estimation}
The efficiency of deep neural networks has been shown in audio/video-related tasks that are challenging for traditional methods\cite{virtanen2018computational,gharib2018acoustic,hinton2012deep,evers2016acoustic,sterling2018isnn}.
%For a more detailed review, please refer to 
%-- here
%we highlight a few that are closely related to ours.
Hershey et al.\ showed that it is feasible to use CNNs for large-scale audio classification problems~\cite{hershey2017cnn}.
Many deep neural networks require a large amount of training data.
Salamon et al.\ used data augmentation to improve environmental sound classification~\cite{salamon2017deep}.
Similarly, Bryan estimates the $T_{60}$ and the direct-to-reverberant ratio (DRR)  from a single speech recording via augmented datasets~\cite{bryan2019}. Tang et al.\ trained CRNN
models purely based on synthetic spatial IRs that generalize to real-world recordings~\cite{tang2019regression,tang2019improving,tang2019lowfreq}.
We strategically design an augmentation scheme to address the challenge of equalization's dependence on both IRs and speaker voice profiles,
which is fully complimentary to all prior data-driven methods. % \zhy{mention whether our augmentation is complimentary or different from prior methods?}

\begin{figure}[tb]
\centering 
\includegraphics[width=0.8\linewidth]{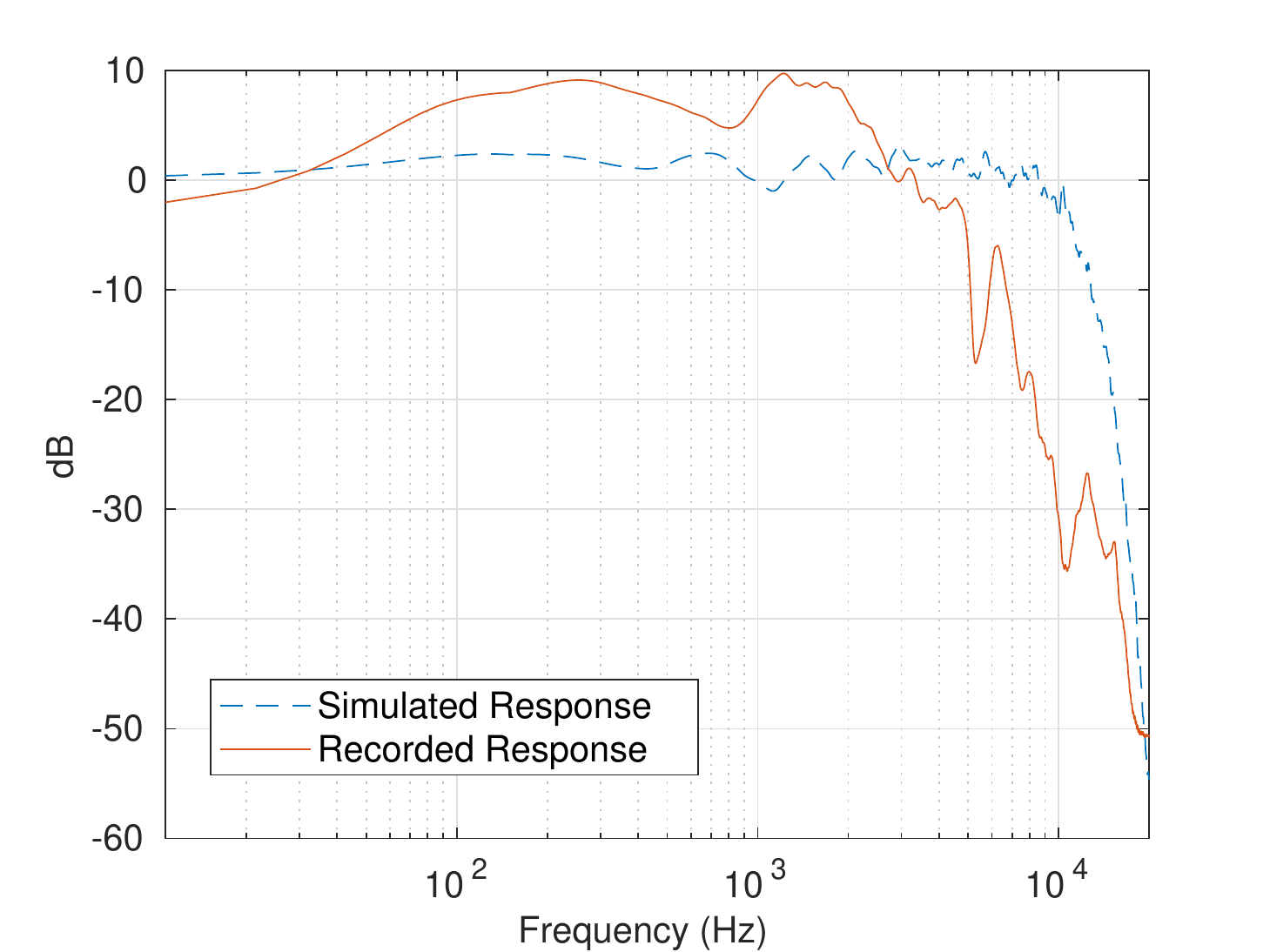}
\caption{The simulated and recorded frequency response in the same room at a sample rate of 44.1kHz is shown. 
Note that the recorded response has noticeable peaks and notches compared with the relatively flat simulated response. 
This is mainly caused by room equalization.
Missing proper room equalization leads to discrepancies in audio quality and overall room acoustics.
}
\label{fig:freq_comp}
\end{figure}

%\paragraph{Inverse Material Modeling}
Acoustic simulators require a set of well-defined material properties. 
%Round robin tests using simulation software show that given accurate enough acoustic parameters, the simulation can yield results that match closely with real-world measurements~\cite{bork2000comparison}. 
The material absorption coefficient is one of the most important parameters~\cite{bork2000comparison}, ranging from 0 (total reflection) to 1 (total absorption). 
% A material's acoustic properties are correlated with its visual appearance to some extent. For example, a carpet is usually more absorptive for sound than a glass is. This audio-visual correlation enables rough material estimation from visual cues~\cite{schissler2017acoustic}. However, despite a non-zero material recognition error, visual information alone does not accurately capture the acoustic property of materials.
% Prior work shows that 7D (source-listener 3D locations and time) acoustic fields in an environment can be effectively compressed into 6D time-invariant fields using only four selected scalar acoustic metrics with low reconstruction errors~\cite{raghuvanshi2014parametric}. This indicates that certain acoustic metrics can be used to guide the modeling of acoustic materials. 
When a reference IR is available, it is straightforward to adjust room materials to match the energy decay of the simulated IR to the reference IR~\cite{Li:2018:360audio}. 
Similarly, Ren et al.\ optimized linear modal analysis parameters to match the given recordings~\cite{ren2013example}. A probabilistic damping model for audio-material reconstruction has been presented for VR applications~\cite{sterling2019audio}.
% In the absence of such a reference due to the difficulty in acquisition, we choose to use the reverberation time~\cite{sabine1953room} as our guideline to inversely drive our material modeling process. \todo{probably define $T_{60}$ here?}
Unlike all previous methods which require a clean IR recording for accurate estimation and optimization of boundary materials, we infer typical material parameters including $T_{60}$ values and equalization from raw speech signals using a CNN-based model.
%\zhy{This sentence should go to contribution paragraph in section 1?} 

\begin{figure*}[tb]
\centering 
\includegraphics[width=\linewidth]{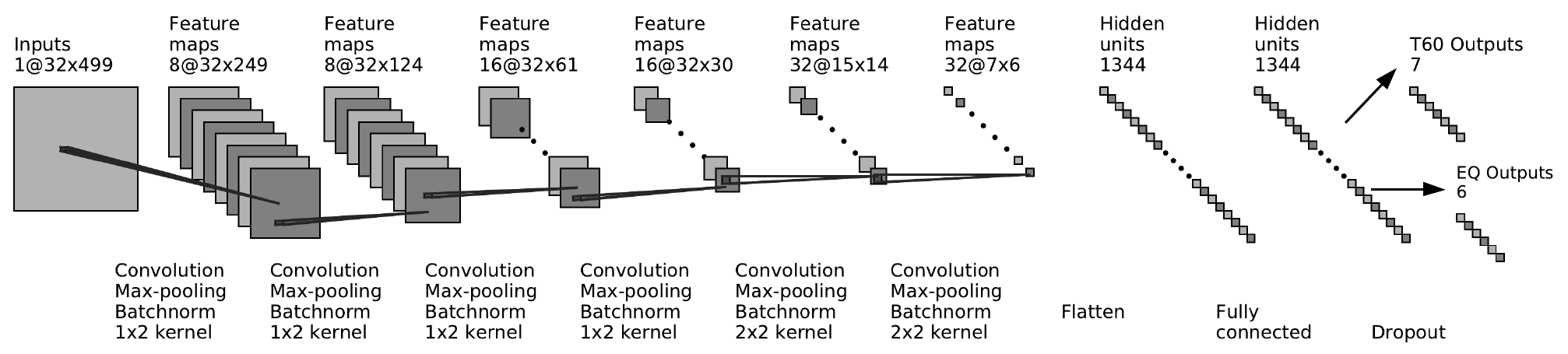}
\caption{Network architecture for $T_{60}$ and EQ prediction. Two models are trained for $T_{60}$ and EQ, which have the same components except the output layers have different dimensions customized for the octave bands they use. }
\label{fig:architecture}
\end{figure*}

\begin{figure}[b]
\centering 
\includegraphics[width=\linewidth]{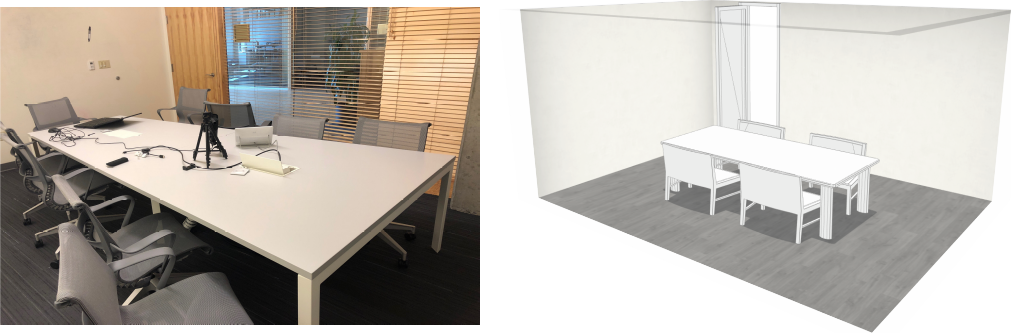}
\caption{We use an off-the-shelf app called MagicPlan to generate geometry proxy. Input: a real-world room (left); Output: the captured 3D model of the room (right) without high-level details, which is used by the runtime geometric acoustic simulator. 
%{\color{red} Shall we move this towards the end of the paper when we talk about runtime, to not make people mistake this as our work?}
% {I don't think so. I will edit the caption to make it clear this is not part of our method. It's just part of the pipeline.}
}
\label{fig:magicplan}
\end{figure}

\begin{figure*}[tb]
\centering 
    \begin{subfigure}[b]{\linewidth}
        \centering
        \includegraphics[width=\linewidth]{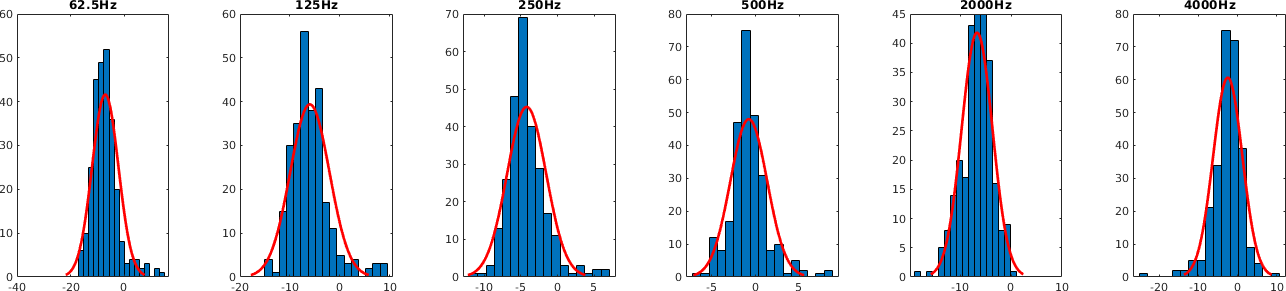}
        \caption{MIT IR survey equalization distribution by sub-band.}
        \label{fig:survey_distribution} 
    \end{subfigure}
    \begin{subfigure}[b]{0.33\linewidth}
        \centering
        \includegraphics[width=\linewidth]{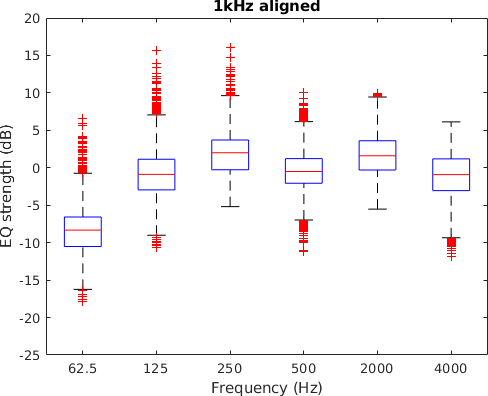}
        \caption{Original synthetic IR equalization.}
        \label{fig:before_aug} 
    \end{subfigure}
    \hfill
    \begin{subfigure}[b]{0.33\linewidth}
        \centering
        \includegraphics[width=\linewidth]{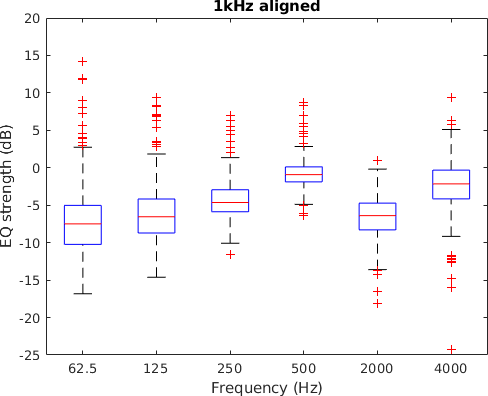}
        \caption{Target (MIT) IR equalization.}
        \label{fig:before_aug} 
    \end{subfigure}
    \hfill
    \begin{subfigure}[b]{0.33\linewidth}
        \centering
        \includegraphics[width=\linewidth]{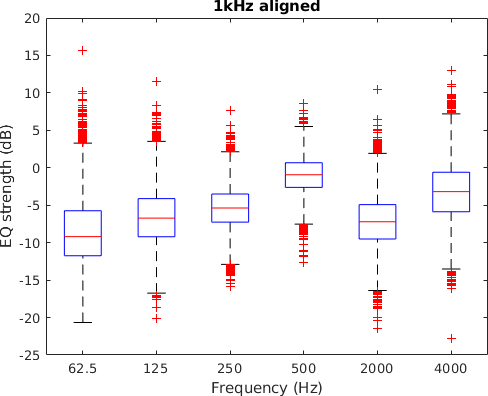}
        \caption{Augmented synthetic IR equalization.}
        \label{fig:after_aug} 
    \end{subfigure}
     \caption{Equalization augmentation. 
     The 1000Hz sub-band is used as reference and has unit gain. 
     We fit normal distributions (red bell curves shown in (a)) to describe the EQ gains of MIT IRs. 
     We then apply EQs sampled from these distributions to our training set distribution in (b). 
     We observe that the augmented EQ distribution in (d) becomes more similar to the target distribution in (c).
      \label{fig:augmentation} 
      }
\end{figure*}

%\todo{inverse optimization, how previous method uses the wrong objective}
Analytical gradients can significantly accelerate the optimization process. 
With similar optimization objectives, it was shown that additional gradient information can boost the speed by a factor of over ten times~\cite{Li:2018:360audio,schissler2017acoustic}. 
The speed gain shown by Li et al.~\cite{Li:2018:360audio} is impressive, and we further improve the accuracy and speed of the formulation. 
%\zhy{It may look bad to be negative about prior work, even though you are the authors... We can just be positive about this work?} Good catch.
More specifically, the original objective function evaluated energy decay relative to the first ray received (the direct sound if there were no obstacles). 
However, energy estimates can be noisy due to both the oscillatory nature of audio as well as simulator noise. 
Instead, we optimize the slope of the best fit line of ray energies to the desired energy decay (defined by the $T_{60}$), which we found to be more robust.

% \todo{everyone, other missing areas?}

% \subsection{Wave Acoustics}

% \subsection{Geometric Acoustics}

% \subsection{Room Impulse Response Acquisition}
% \subsubsection{RIR Measurement}
% \subsubsection{Blind Estimation}
% \subsection{Synthetic Sound for AR/VR}

%%%%%%%%%%%%%%%%%%%%%%%%%%%%%%%%%%%%%%%%%%%%%%%%%%%%%%%%%%%%%%%%%%%%%%%%%%%%%%%%
%%%%%%%%%%%%%%%%%%%%%%%%%%%%%%%%%%%%%%%%%%%%%%%%%%%%%%%%%%%%%%%%%%%%%%%%%%%%%%%%
%%%%%%%%%%%%%%%%%%%%%%%%%%%%%%%%%%%%%%%%%%%%%%%%%%%%%%%%%%%%%%%%%%%%%%%%%%%%%%%%
\section{Deep Acoustic Analysis: Our Algorithm}
In this section, we overview our proposed method for scene-aware audio rendering. We begin by providing background information, discuss how we capture room geometry, and then proceed with discussing how we estimate the frequency dependent room reverberation and equalization parameters directly from recorded speech. We follow by discussing how we use the estimated acoustic parameters to perform acoustic materials optimization such that we calibrate our virtual acoustic model with real-world recordings.

%\section{Rationale and Overview}
% Before diving into technical details, we start with our motivations for building our system in a two-stage approach including inverse material modeling and room equalization compensation, which are characteristic features of an enclosed acoustic environment.

% \subsection{Room Equalization Compensation}
    % Most modern real-time sound engines employ geometric sound propagation due to its efficiency, compared with wave based solutions. However, acoustic wave phenomena such as sound diffraction and room modes are usually missing from the pipeline. Though there are object based pre-computation methods that can incorporate sound diffraction effects into geometric methods~\cite{rungta2018diffraction}, pre-computing room modes using wave based solvers can still be overwhelming for interactive applications. Room modes occur as standing waves appear in a room for steady-state signals, which is mainly determined the room layout. At room resonance frequencies, the frequency response of the room tends to have peaks and notches. Room equalization is mostly obvious at low frequencies, and is distinct for each acoustic environment~\cite{cecchi2018room}, which makes it impractical to use a generic one for all rooms. 
 
%\section{Equalization Prediction and Optimization}

% \todo{introduce $T_{60}$?}

\begin{table}[tb]
\centering
    \caption{Notation and symbols used throughout the paper.}
    \begin{tabular}{ll}
    \toprule
    % \begin{tabular}{p{0.06\columnwidth}p{0.9\columnwidth}}
        $T_{60}$ & Reverberation time for sound energy to drop by 60dB. \\
        $t$ & Sound arrival time. \\
        $\rho$ & Frequency dependent sound absorption coefficient. \\
        $e_{j}$ & Energy carried by a sound path $j$. \\
        $\beta$ & Air absorption coefficient.\\
        $m$ & Slope of the energy curve envelope.\\
        \bottomrule
    \label{tab:notation}
    \end{tabular}
\end{table}

%%%%%%%%%%%%%%%%%%%%%%%%%%%%%%%%%%%%%%%%%%%%%%%%%%%%%%%%%%%%%%%%%%%%%%%%%%%%%%%%
%%%%%%%%%%%%%%%%%%%%%%%%%%%%%%%%%%%%%%%%%%%%%%%%%%%%%%%%%%%%%%%%%%%%%%%%%%%%%%%%
\subsection{Background}
To explain the motivation of our approach, we briefly elaborate on the most difficult parts of previous approaches, upon which our method improves.
Previous methods require an impulse response of the environment to estimate acoustic properties~\cite{Li:2018:360audio,schissler2017acoustic}.
Recording an impulse response is a non-trivial task. 
The most reliable methods involve playing and recording Golay codes \cite{foster1986impulse} or sine sweeps \cite{farina2000simultaneous}, which both play loud and intrusive audio signals.
Also required are a fairly high-quality speaker and microphone with constant frequency response, small harmonic distortion and little crosstalk. 
The speaker and microphone should be acoustically separated from surfaces, i.e., they shouldn't be placed directly on tables (else surface vibrations could contaminate the signal). 
Clock drift between the source and microphone must be accounted for~\cite{Bryan2010impulse}. 
Alternatively, balloon pops or hand claps have been proposed for easier IR estimation, but require significant post-processing and still are very obtrusive~\cite{abel2010estimating, seetharaman2012hand}.
In short, correctly recording an IR is not easy, and makes it challenging to add audio in scenarios such as augmented reality, where the environment is not known beforehand and estimation must be done interactively to preserve immersion.

Geometric acoustics is a high-frequency approximation to the wave equation. 
It is a fast method, but assumes that wavelengths are small compared to objects in the scene, while ignoring pressure effects~\cite{savioja2015overviewGA}. 
It misses several important wave effects such as diffraction and room resonance.
Diffraction occurs when sound paths bend around objects that are of similar size to the wavelength. Resonance is a pressure effect that happens when certain wavelengths are either reinforced or diminished by the room geometry: certain wavelengths create peaks or troughs in the frequency spectrum based on the positive or negative interference they create~\cite{cox2004room}. 

We model these effects with a linear finite impulse response (FIR) equalization filter~\cite{schafer1989discrete}.
We compute the discrete Fourier transform on the recorded IR
%$  \tilde{\mathsf{H}}(\omega) = \mathcal{F}[\tilde{H}(t)]$ 
over all frequencies, following~\cite{Li:2018:360audio}.
Instead of filtering directly in the frequency domain,
we design a linear phase EQ filter with 32ms delay to compactly represent this filter at 7 octave bin locations.
% \zhy{Explain how we extract EQ here. The ``compact" representation in the next sentence means picking octave bands on the response curve to represent it. A linear phase EQ filter is designed based on these sample points with a 32ms delay.}
We then blindly estimate this compact representation of the frequency spectrum of the impulse response as discrete frequency gains, without specific knowledge of the input sound or room geometry. 
This is a challenging estimation task. Since the convolution of two signals (the IR and the input sound) is equivalent to multiplication in the frequency domain, estimating the frequency response of the IR is equivalent to estimating one multiplicative factor of a number without constraining the other. 
We are relying on this approach to recognize a compact representation of the frequency response magnitude in different environments.

%%%%%%%%%%%%%%%%%%%%%%%%%%%%%%%%%%%%%%%%%%%%%%%%%%%%%%%%%%%%%%%%%%%%%%%%%%%%%%%%
%%%%%%%%%%%%%%%%%%%%%%%%%%%%%%%%%%%%%%%%%%%%%%%%%%%%%%%%%%%%%%%%%%%%%%%%%%%%%%%%
\subsection{Geometry Reconstruction}
Given the background, we begin by first estimating the room geometry. In our experiments, we utilize the ARKit-based iOS app MagicPlan\footnote{https://www.magicplan.app/} to acquire the basic room geometry. 
A sample reconstruction is shown in Figure~\ref{fig:magicplan}.
With computer vision research evolving rapidly, we believe constructing geometry proxies from video input will become even more robust and easily accessible~\cite{Zhi_2019_CVPR,Bloesch_2018_CVPR}.

\subsection{Learning Reverberation and Equalization}
\label{sec:learning}
We use a convolutional neural network (Figure~\ref{fig:architecture}) to predict room equalization and reverberation time $(T_{60})$ directly from a speech recording. 
Training requires a large number of speech recordings with known $T_{60}$ and room equalization. 
The standard practice is to generate speech recordings from known real-world or synthetic IRs~\cite{Kim2017Generation, Doulaty2017Automatic}. 
Unfortunately, large scale IR datasets do not currently exist due to the difficulty of IR measurement; most publicly available IR datasets have fewer than 1000 IR recordings. Synthetic IRs are easy to obtain and can be used, but again lack wave-based effects as well as other simulation deficiencies. Recent work has addressed this issue by combining real-word IR measurements with augmentation to increase the diversity of existing real-world datasets~\cite{bryan2019}. This work, however, only addresses $T_{60}$ and DRR augmentation, and lacks a method to augment the frequency-equalization of existing IRs. To address this, we propose a method to do this in Section~\ref{sec:dataaugmentation}. Beforehand, however, we discuss our neural network estimation method for estimating both $T_{60}$ and equalization.

%%%%%%%%%%%%%%%%%%%%%%%%%%%%%%%%%%%%%%%%%%%%%%%%%%%%%%%%%%%%%%%%%%%%%%%%%%%%%%%%
\subsubsection{Octave-Based Prediction}
\label{sec:octave}
Most prior work takes the full-frequency range as input for prediction.
For example, one closely related work~\cite{bryan2019} only predicts one  $T_{60}$ value for the entire frequency range (full-band).
However, sound propagates and interacts with materials differently at different frequencies. 
To this end, we define our learning targets over several octaves. 
%Because the acoustic simulation involves frequency dependent parameters, we also mark our targets in octave bands. 
Specifically, we calculate $T_{60}$ at 7 sub-bands centered at $\{125, 250, 500, 1000, 2000, 4000, 8000\}$Hz. %\zhy{should we explain how to compute them?}.
% \todo{why are we missing 62.5Hz??}\zhy{Error is too high due to inaccurate labeling (even for test set), which is not the fail case of our method, but $T_{60}$ calculation on low SNR sub-band.}
%Consequently, we have both the $T_{60}$ and the equalization as vectors of length 7.
%As for equalization, because it is a relative metric specifying the gain contrast among frequency bands, we set the gain at 1000Hz to be $0dB$ as a reference, and sample frequency gains at $\{62.5, 125, 250, 500, 2000, 4000, 8000\}$Hz.
% The choice of octave band filter is important, as we will show next how it would affect the equalization augmentation and in turn the training accuracy. \todo{where do we show this?}
% \todo{should be in the equalization augment section. Mention the early method that failed.}\zhy{The failed method should not have been considered in the first place since it's not the standard practice. I suggest we skip this.}
We found prediction of $T_{60}$ at the 62.5Hz band to be unreliable due to low signal-to-noise ratio (SNR). During material optimization, we set the 62.5Hz $T_{60}$ value to the 125Hz value. Our frequency equalization estimation is done at 6 octave bands centered at $\{62.5, 125, 250, 500, 2000, 4000\}$Hz. As we describe in \S
\ref{sec:dataaugmentation}, we compute equalization relative to the 1kHz band, so we do not estimate it. When applying our equalization filter, we set bands greater than or equal to 8kHz to $-50$dB. Given our target sampling rate of 16kHz and the limited content of speech in higher octaves, this did not affect our estimation.

%This octave-based prediction also aligns with the geometric acoustic simulator since the absorption coefficients are also octave-based. 
%By training and predicting on sub-band frequencies, the mapping from $T_{60}$ prediction to our optimized simulator is seamless. 

%%%%%%%%%%%%%%%%%%%%%%%%%%%%%%%%%%%%%%%%%%%%%%%%%%%%%%%%%%%%%%%%%%%%%%%%%%%%%%%%
\subsubsection{Data Augmentation} \label{sec:dataaugmentation}
We use the following datasets as the basis for our training and augmentation. 
\begin{itemize}
    \item ACE Challenge: 70 IRs and noise audio~\cite{eaton2016estimation};
    \item MIT IR Survey: 271 IRs~\cite{traer2016statistics};
    \item DAPS dataset: 4.5 hours of 20 speakers' speech (10 males and 10 females)~\cite{mysore2014can}.
\end{itemize}

% We design our augmentation based on Bryan~\cite{bryan2019}. 
%For DRR and reverberation augmentation, we follow~\cite{bryan2019} and construct a \todo{synthetic IR dataset? need more details.}
% We randomly convolve an IR either from synthetic IRs or MIT survey IRs with an 8-second speech segment, and add noise with signal-to-noise ratios (SNRs) uniformly distributed between $-5\sim 20dB $. 
% The mixed audio becomes our output dataset, which is further split into training (56,500 samples), validation (19,400 samples), and test (18,450 samples) sets.
% \paragraph{Reverberation Augmentation.} 
%We first collect 70 multi-channel IRs and noise audios from the ACE Challenge~\cite{eaton2016estimation}. 
% Because these IRs have an unbalanced distribution in terms of $T_{60}$ and direct-to-reverb ratio (DRR), which increases the risk of overfitting, first we augment IRs by applying our $T_{60}$ and DRR augmentation procedure in sequence. 
% We specify $T_{60}$ values to be uniformly distributed between $0.1\sim 1.5$ seconds and DRR values to be uniformly distributed between $-6\sim 18dB $, resulting in 7000 (100x) synthetic IRs with a balanced distribution \todo{\cite{bryan2019}}. 
% We also collect 271 IRs from the MIT IR survey~\cite{traer2016statistics} to use as an independent test set. 
% Then we collect clean speech from the Device and Produced Speech (DAPS) dataset~\cite{mysore2014can}, which consists of 4.5 hours of 20 speakers' speech (10 males and 10 females). 
%\paragraph{Equalization Augmentation.} 

First, we use the method in~\cite{bryan2019} to expand the $T_{60}$ and direct-to-reverberant ratio (DRR) range of the 70 ACE IRs, resulting in 7000 synthetic IRs with a balanced $T_{60}$ distribution between 0.1--1.5 seconds. The ground truth $T_{60}$ estimates can be computed directly from IRs can be computed is a variety of ways. We follow the methodology of Karjalainen et  al.~\cite{Karjalainen2001estimation} when computing the $T_{60}$ from real IRs with a measurable noise floor. This method was found to be the most robust estimator when computing the $T_{60}$ from real IRs in recent work~\cite{eaton2016estimation}. %After we apply the method of Bryan~\cite{bryan2019} to augment our dataset, however, any noise floor effects are removed, allowing us to simply use linear regression in the log-domain for $T_{60}$ labeling. 
The final composition of our dataset is listed in Table~\ref{tab:dataset}.

While we know the common range of real-world $T_{60}$ values, there is limited literature giving statistics about room equalization. 
Therefore, we analyzed the equalization range and distribution of the 271 MIT survey IRs as a guidance for data augmentation. 
The equalization of frequency bands is computed relative to the 1kHz octave. This is a common practice~\cite{valimaki2016all}, unless expensive equipment is used to obtain calibrated acoustic pressure readings. 

For our equalization augmentation procedure, we first fit a normal distribution (mean and standard deviation) to each sub-band amplitude of the MIT IR dataset as shown in Figure~\ref{fig:augmentation}. Given this set of parametric model estimates, we iterate through our training and validation IRs.  For each IR, we extract its original EQ. We then randomly sample a target EQ according to our fit models (independently per frequency band), calculate the distance between the source and target EQ, and then design an FIR filter to compensate for the difference. For simplicity, we use the window method for FIR filter design~\cite{SASPWEB2011}. Note, we do not require  a perfect filter design method. We simply need a procedure to increase the diversity of our data. Also note, we intentionally sample our augmented IRs to have a larger variance than the recorded IRs to further increase the variety of our training data.

\begin{table}[tb]
\centering
\caption{Dataset composition. The training set and validation set are based on synthetic IRs and the test set is based on real IRs to guarantee model generalization. Clean speech files are also divided in a way that speakers (``f1" for female speaker 1; ``m10" for male speaker 10) in each dataset partition are different, to avoid the model learning the speaker's voice signature. Audio files are generated at a sample rate of 16kHz, which is sufficient to cover the human voice's frequency range.}
% \small
\label{tab:dataset}
\begin{tabular}{cccc}
\toprule
Partition  & Noise  & Clean Speech  & IR      \\\hline
\thead{Training set\\(size: 56.5k)}    & ACE ambient & f5$\sim$f10, m5$\sim$m10 & \thead{Synthetic IR\\(size: 4.5k)} \\
\thead{Validation set\\(size: 19.5k)} & ACE ambient & f3, f4, m3, m4   & \thead{Synthetic IR\\(size: 1k)}   \\
\thead{Test set\\(size: 18.5k)}  & ACE ambient & f1, f2, m1, m2             & \thead{MIT survey IR\\(size: 271)} \\\bottomrule
\end{tabular}
\end{table}

%\paragraph{Feature Extraction.} 
We compute the log Mel-frequency spectrogram for each four second audio clip, which is commonly used for speech-related tasks~\cite{Chen_2018_ECCV,eskimez2018front}. 
We use a Hann window of size 256 with 50\% overlap during computation of the short-time Fourier transform (STFT) for our 16kHz samples. 
Then we use 32 Mel-scale bands and area normalization for Mel-frequency warping~\cite{stevens1937scale}. 
The spectrogram power is computed in decibels. This extraction process yields a 32 x 499 (frequency x time domain) matrix feature representation. 
All feature matrices are normalized by the mean and standard deviation of the training set.

%%%%%%%%%%%%%%%%%%%%%%%%%%%%%%%%%%%%%%%%%%%%%%%%%%%%%%%%%%%%%%%%%%%%%%%%%%%%%%%%
\subsubsection{Network Architecture and Training}
\label{sec:architecture}
%Overall, we follow the common network design in 2D visual learning problems and use a convolutional neural network (CNN). 
% \todo{So this is essentially Nick's model design?}
% \zhy{True. Except the output layer is different. Mostly the same because it works.}
We propose using a network architecture differing only in the final layer for both $T_{60}$ and room equalization estimation. 
Six 2D convolutional layers are used sequentially to reduce both the time and frequency resolution of features until they have approximately the same dimension.
Each \texttt{conv} layer is immediately followed by a rectified linear unit (\texttt{ReLU})~\cite{nair2010rectified} activation function, 2D max pooling, and batch normalization. 
The output from \texttt{conv} layers is flattened to a 1D vector and connected to a fully connected layer of 64 units, at a dropout rate of 50\% to lower the risk of overfitting. 
The final output layer has 7 fully connected units to predict a vector of length 7 for $T_{60}$ or 6 fully connected units to predict a vector of length 6 for frequency equalization. This network architecture is inspired by Bryan~\cite{bryan2019}, where it was used to predict full-band $T_{60}$. We updated the output layer to predict the more challenging sub-band $T_{60}$, and also discovered that the same architecture predicts equalization well.

% \todo{I thought our modulation prediction network is much more complicated than $T_{60}$?}
% {from Zhenyu: BTW, I ended up using the same simple network for both models because I found the addition of recurrent layers for modulation only helps reduce validation error, but has similar/slightly higher test error compared with the simple CNN one...}

For training the network, we use the mean square error (MSE) with the ADAM optimizer~\cite{kingma2014adam} in Keras~\cite{chollet2015keras}. 
The maximum number of epochs is 500 with an early stopping mechanism. 
We choose the model with the lowest validation error for further evaluation on the test set. 
Our model architecture is shown in Figure~\ref{fig:architecture}.
% \todo{better visualization? redo Figure~\ref{fig:architecture}?}

%%%%%%%%%%%%%%%%%%%%%%%%%%%%%%%%%%%%%%%%%%%%%%%%%%%%%%%%%%%%%%%%%%%%%%%%%%%%%%%%
\subsection{Acoustic Material Optimization}
% In geometric sound propagation, sound energy are released from the sound source in the form of directed rays that carry some portion of the total energy. The IR essentially describes when each ray arrives at the listener and how much energy each ray still carries. The loss of energy is attributed to attenuation in the air, and absorption by surface materials. The attenuation in air is only related to a constant air absorption coefficient ($dB/m$) and the total path length a ray has travelled. Most indoor sound energy loss comes from material absorption, so the energy decay of an IR heavily depends on that.
Our goal is to optimize the material absorption coefficients at the same octave bands as our $T_{60}$ estimator in \S~\ref{sec:octave} of a set of room materials to match the sub-band $T_{60}$ of the simulated sound with the target predicted in \S~\ref{sec:learning}.

\paragraph{Ray Energy.} We borrow notation from~\cite{Li:2018:360audio}. Briefly, a geometric acoustic simulator generates a set of sound paths, each of which carries an amount of sound energy. Each material $m_i$ in a scene is described by a frequency dependent absorption coefficient, $\rho_i$. A path leaving the source is reflected by a set of materials before it reaches the listener. The energy fraction that is received by the listener along path $j$ is
\begin{equation}
    e_j = \beta_j \prod_{k=1}^{N_j} \rho_{m^k},
\end{equation}
where $m^k$ is the material the path intersects on the $k^{th}$ bounce, $N_j$ is the number of surface reflections for path $j$, and $\beta_j$ accounts for air absorption (dependent on the total length of the path). Our goal is to optimize the set of absorption coefficients $\rho_i$ to match the energy distribution of the paths $e_j$ to that of the environment's IR. Again similar to~\cite{Li:2018:360audio}, we assume the energy decrease of the IR follows an exponential curve, which is a linear decay in dB space. The slope of this decay line in dB space is $m' = -60 / T_{60}$.

% \begin{figure}[tb]
% \centering 
% \includegraphics[width=\linewidth]{pictures/placeholder_flat.pdf}
% \caption{Better Objective function illustration -- to show what is different in~\cite{Li:2018:360audio} and ours.}
% \label{fig:better_objective}
% \end{figure}

\paragraph{Objective Function.} 
We propose the following objective function:
\begin{equation}
J(\rho) = (m - m')^2
\end{equation}
where $m$ is the best fit line of the ray energies on a decibel scale:
\begin{equation}
m = \frac{n \sum_{i=0}^n t_i y_i - \sum_{i=0}^n t_i \sum_{i=0}^n y_i} {n \sum_{i=0}^n t_i^2 - \left( \sum_{i=0}^n t_i \right)^2}, 
\end{equation}
with $y_i = 10 \text{log}_{10}(e_i)$, which we found to be more robust than previous methods. %~\cite{Li:2018:360audio}.
Specifically, in comparison with Equation (3) in~\cite{Li:2018:360audio}, we see that Li et al.\ tried to match the slope of the energies relative to $e_0$, forcing $e_0$ to be at the origin on a dB scale. 
%\zhy{Move this to Optimization paragraph in section 4, since we have presented the actual comparison results there?}
%\todo{I think this is to explain the math, not the actual results/slope.}
However, we only care about the energy decrease, and not the absolute scale of the values from the simulator.
We found that allowing the absolute scale to move and only optimizing the slope of the best fit line produces a better match to the target $T_{60}$. 
% Figure~\ref{fig:better_objective} illustrates the difference in the objectives and how ours better models the optimization. \zhy{how do we want to change this sentence?}

We minimize $J$ using the L-BFGS-B algorithm~\cite{Zhu:1997:LBFGS}.
The gradient of $J$ is given by
\begin{equation}
\frac{\partial J}{\partial \rho_j} = 2 (m - m') \frac{n t_i - \sum_{i=0}^n t_i}{n \sum_{i=0}^n t_i^2 - \left( \sum_{i=0}^n t_i \right)^2} \frac{10}{\text{ln}(10) e_i} \frac{\partial e_i}{\partial \rho_j}
\end{equation}

%Key: improved objective function

\begin{figure}[tb]
\centering 
    \begin{subfigure}[b]{0.49\linewidth}
        \centering
        \includegraphics[width=\linewidth]{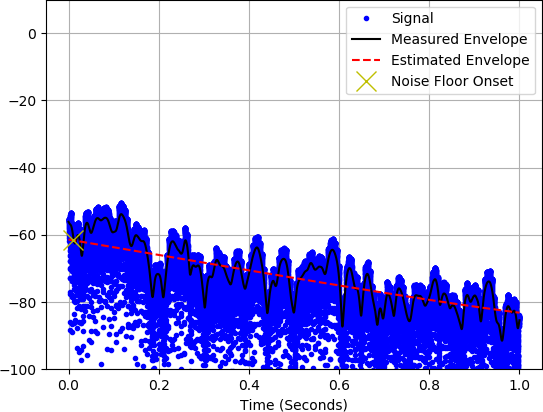}
        \caption{125Hz sub-band.}
    \end{subfigure}
    \begin{subfigure}[b]{0.49\linewidth}
        \centering
        \includegraphics[width=\linewidth]{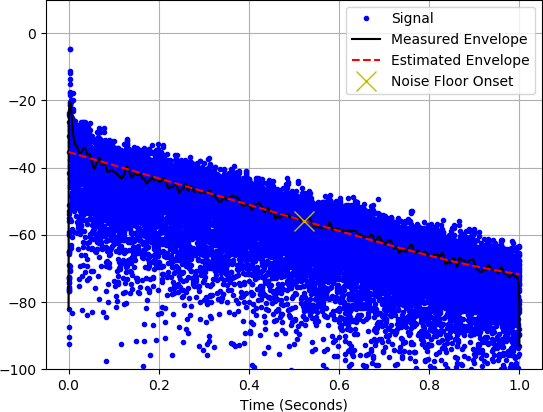}
        \caption{8000Hz sub-band.}
    \end{subfigure}
     \caption{Evaluating $T_{60}$ from signal envelope on low and high frequency bands of the same IR. 
     Note that the SNR in the low frequency band is lower than the high frequency band. %, limited by recording conditions. 
     This makes $T_{60}$ evaluation for low frequency bands less reliable, which partly explains the larger test error in low frequency sub-bands.}
 \label{fig:lowfreq_t60} 
\end{figure}

%%%%%%%%%%%%%%%%%%%%%%%%%%%%%%%%%%%%%%%%%%%%%%%%%%%%%%%%%%%%%%%%%%%%%%%%%%%%%%%%
%%%%%%%%%%%%%%%%%%%%%%%%%%%%%%%%%%%%%%%%%%%%%%%%%%%%%%%%%%%%%%%%%%%%%%%%%%%%%%%%
%%%%%%%%%%%%%%%%%%%%%%%%%%%%%%%%%%%%%%%%%%%%%%%%%%%%%%%%%%%%%%%%%%%%%%%%%%%%%%%%
\section{Analysis and Applications}
% \section{Analysis and Applications}
% \zhy{how about ``Results and Analysis" as section title?}

\subsection{Analysis}
\paragraph{Speed.} 
%W-2123
We implement our system on an Intel Xeon(R) CPU @3.60GHz and an NVIDIA GTX 1080 Ti GPU. 
Our neural network inference runs at 222 frames per second (FPS) on 4-second sliding windows of audio due to the compact design (only 18K trainable parameters).
Optimization runs twice as fast with our improved objective function.
The sound rendering is based on the real-time geometric bi-directional sound path tracing from Cao et al.~\cite{cao2017bidirectional}.
%\zhy{Shall we move hardware info to Analysis section (or a ``real-time" subsection)? Not only CNN uses these resources.}

{
\renewcommand{\arraystretch}{1.5}
\begin{table*}[t]
\centering
\caption{Benchmark results for acoustic matching. These real-world rooms are of different sizes and shapes, and contain a wide variety of acoustic materials such as brick, carpet, glass, metal, wood, plastic, etc., which make the problem acoustically challenging.
We compare our method with \cite{Li:2018:360audio}. 
Our method does not require a reference IR and still obtains similar $T_{60}$ and EQ errors in most scenes compared with their method.
We also achieve faster optimization speed. 
Note that the input audio to our method is already noisy and reverberant, whereas \cite{Li:2018:360audio} requires clean IR recording. All IR plots in the table have the same time and amplitude scale.}
\label{tab:benchmark}
\begin{tabular}{C{0.1\textwidth}L{0.15\textwidth}L{0.15\textwidth}L{0.15\textwidth}L{0.15\textwidth}L{0.15\textwidth}}
% {C{0.6in}L{1in}L{1in}L{1in}L{1in}L{1in}}
\toprule
                %   & Davis & 301   & 501   & 620   & 750   \\
Benchmark Scene & \tabfig{1.1600}{benchmarks/benchmark_annotateDavis_cropped.png}      &  \tabfig{1.1600}{benchmarks/benchmark_annotate301_cropped.png}     &    \tabfig{1.1600}{benchmarks/benchmark_annotate501_cropped.png}   &    \tabfig{1.1600}{benchmarks/benchmark_annotate620_cropped.png}   &   \tabfig{1.1600}{benchmarks/benchmark_annotate750_cropped.png}    \\
Size ($m^3$)           & 1100 (irregular) & 1428 (12$\times$17$\times$7) & 990 (11$\times$15$\times$6) & 72 (4$\times$6$\times$3) & 352 (11$\times$8$\times$4) \\
\# Main planes              & 6     & 6     & 6     & 11     & 6     \\
\thead{Groundtruth IR\\(dB scale)} & \scalebox{1.0}[2.50001]{\tabfig{1.1600}{benchmarks/davisref.png}}  & \scalebox{1.0}[2.50001]{\tabfig{1.1600}{benchmarks/301ref.png}}   & \scalebox{1.0}[2.50001]{\tabfig{1.1600}{benchmarks/501ref.png}}   & \scalebox{1.0}[2.50001]{\tabfig{1.1600}{benchmarks/620ref.png}}   & \scalebox{1.0}[2.50001]{\tabfig{1.1600}{benchmarks/750ref.png}}   \\\hline
\thead{Li et al.~\cite{Li:2018:360audio} IR\\(dB scale)}   & \scalebox{1.0}[2.50001]{\tabfig{1.1600}{benchmarks/davisbaseA.png}}  & \scalebox{1.0}[2.50001]{\tabfig{1.1600}{benchmarks/301baseA.png}}  & \scalebox{1.0}[2.50001]{\tabfig{1.1600}{benchmarks/501baseA.png}}  & \scalebox{1.0}[2.50001]{\tabfig{1.1600}{benchmarks/620baseA.png}}  & \scalebox{1.0}[2.50001]{\tabfig{1.1600}{benchmarks/750baseA.png}}  \\
Opt. time (s) & 29 &43 &25 &71 &46 \\
$T_{60}$ error (s) & 0.11 &0.23 &0.08 &0.02 &0.10\\
EQ error (dB) & 1.50 & 2.97 & 8.59 & 3.61 & 7.55 \\\hline
\thead{Ours IR \\(dB scale)}               & \scalebox{1.0}[2.50001]{\tabfig{1.1600}{benchmarks/davispropA.png}}  & \scalebox{1.0}[2.50001]{\tabfig{1.1600}{benchmarks/301propA.png}}  & \scalebox{1.0}[2.50001]{\tabfig{1.1600}{benchmarks/501propA.png}}  & \scalebox{1.0}[2.50001]{\tabfig{1.1600}{benchmarks/620propA.png}}  & \scalebox{1.0}[2.50001]{\tabfig{1.1600}{benchmarks/750propA.png}}\\
Opt. time (s) &13 &13 &14 &31 &20 \\
$T_{60}$ error (s) & 0.14 &0.12 &0.10 &0.04 &0.24\\
EQ error (dB) & 2.26 & 3.86 & 3.97 & 3.46 & 4.62 \\\bottomrule
\end{tabular}
\end{table*}
}

% \subsection{$T_{60}$ Prediction and Matching}
% \zhy{Plan for this subsection: report validation and test errors; analyze possible error sources: Figure~\ref{fig:lowfreq_t60} shows low SNR for low bands so the label from ACE IR itself is inaccurate; mention results from ACE challenge to make the point that sub-band T60 estimation is more difficult than full-band and we achieve results comparable to full-band results on 3rd party real data.}

\paragraph{Sub-band $T_{60}$ prediction.} 
We first evaluate our $T_{60}$ blind estimation model and achieve a mean absolute error (MAE) of 0.23s on the test set (MIT IRs). While the 271 IRs in the test set have a mean $T_{60}$ of 0.49s with a standard deviation (STD) of 0.85s at the 125Hz sub-band, the highest sub-band 8000Hz only has a mean $T_{60}$ of 0.33s with a STD of 0.24s, which reflects a narrow subset within our $T_{60}$ augmentation range.
% \todo{What's the dataset's average T60 values? Like Tim suggested, maybe also list the error in terms of percentages?}
% \zhy{Reporting the average of MIT IR does not tell us its distribution. And display error in terms of percentages is going to hurt us. Because for some outlier, you can get 1000\% error...}
We also notice that the validation MAE on ACE IRs is 0.12s, which indicates our validation set and the test set still come from different distributions.
Another error source is the inaccurate labeling of low-frequency sub-band $T_{60}$ as shown in Figure~\ref{fig:lowfreq_t60}, but we do not filter any outliers in the test set.
In addition, our data is intended to cover frequency ranges up to 8000Hz, but human speech has less energy in high-frequency range~\cite{TITZE2017382}, which results in low signal energy for these sub-bands, making it more difficult for learning. % \zhy{Maybe show higher error in 8k band?}.
% another link on human vocal range
% http://www.bnoack.com/index.html?http&&&www.bnoack.com/audio/speech-level.html

\begin{figure}[tb]
\centering 
\includegraphics[width=\linewidth]{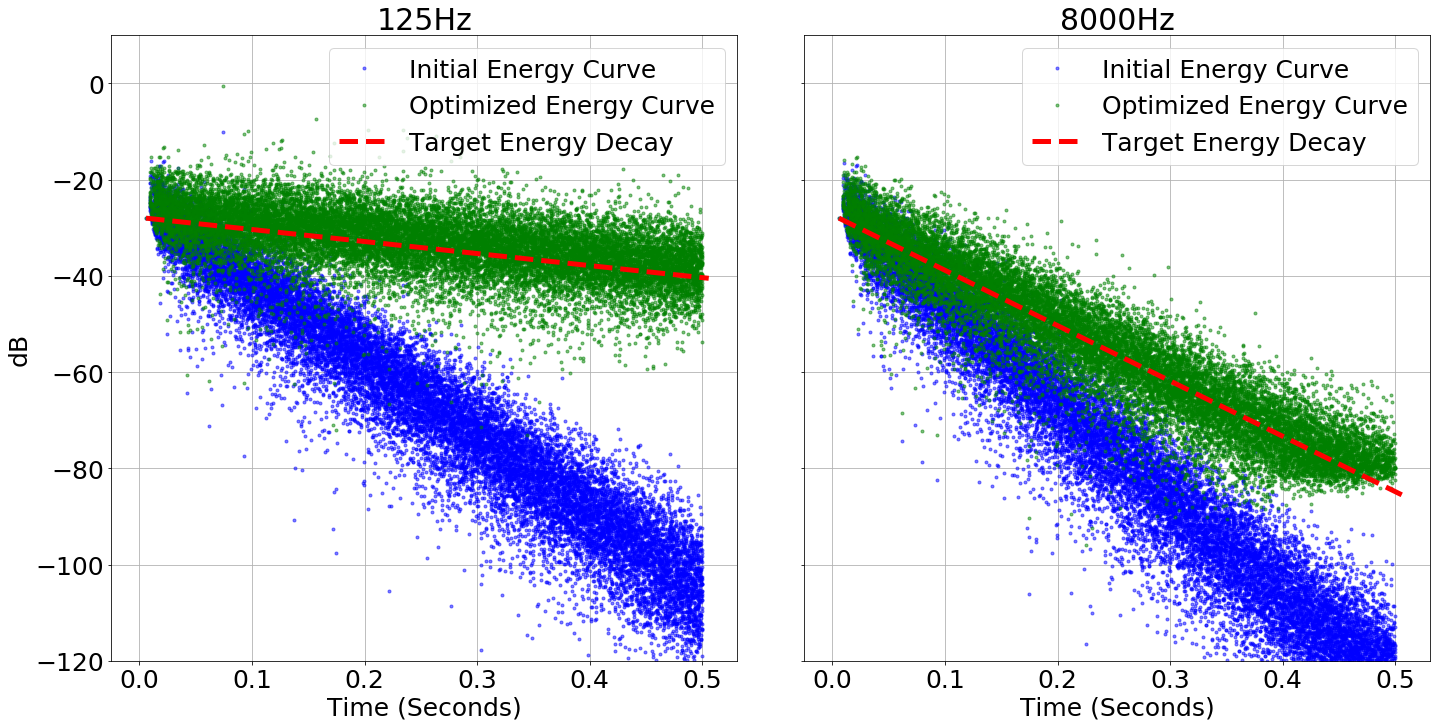}
\caption{Simulated energy curves before and after optimization (with target slope shown).}
\label{fig:optimization_comparison}
\end{figure}

\begin{figure}[tb]
\centering 
\includegraphics[width=0.75\linewidth]{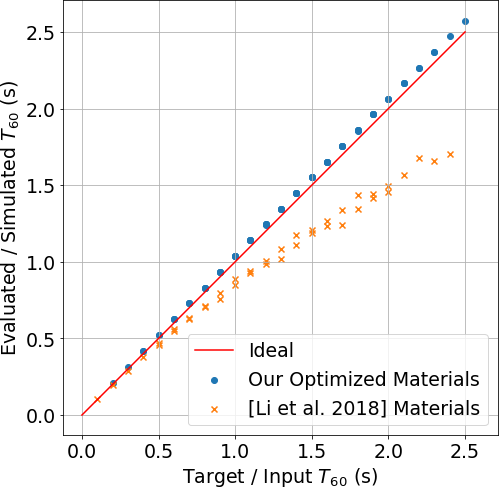}
\caption{Stress test of our optimizer.
We uniformly sample $T_{60}$  between 0.2s and 2.5s and set it to be the target. 
The ideal I/O relationship is a straight line passing the origin with slope $1$.
Our optimization results matches the ideal line much better than prior optimization method.
}
\label{fig:sweep}
\end{figure}

\paragraph{Material Optimization.}
When we optimize the room material absorption coefficients according to the predicted $T_{60}$ of a room, our optimizer efficiently modifies the simulated energy curve to a desired energy decay rate ($T_{60}$) as shown in Figure~\ref{fig:optimization_comparison}. 
We also try fixing the room configuration and set the target $T_{60}$ to values uniformly distributed between 0.2s and 2.5s, and evaluate the $T_{60}$ of the simulated IRs. 
The relationship between the target and output $T_{60}$ is shown in Figure~\ref{fig:sweep}, in which our simulation closely matches the target, demonstrating that our optimization is able to match a wide range of $T_{60}$ values.

To test the real-world performance of our acoustic matching, we recorded ground truth IRs in 5 benchmark scenes, then use the method in~\cite{Li:2018:360audio}, which requires a reference IR, and our method, which does not require an IR, for comparison. Benchmark scenes and results are summarized in Table~\ref{tab:benchmark}.
We apply the EQ filter to the simulated IR as a last step. Overall, we obtain a prediction MAE of 3.42dB on our test set, whereas before augmentation, the MAE was 4.72dB under the same training condition, which confirms the effectiveness of our EQ augmentation.
% Their sub-band MAEs are compared in 
% Figure~\ref{fig:augment}, which suggests our augmentation is effective in reducing the EQ prediction error. 
The perceptual impact of the EQ filter step is evaluated in \S\ref{sec:study}.

% \begin{figure}[htbp]
% \centering 
% \includegraphics[width=0.8\linewidth]{pictures/noise.png}
% \caption{$T_{60}$ prediction error as SNR decreases. {\color{red} make y axis a relative scale, or report the true $T_{60}$ in the caption.}}
% \label{fig:noise}
% \end{figure}

% \paragraph{Noise Resilience.}
% Our model still produce reasonable prediction with noisy background audio. \todo{for Zhenyu to run two audios, one clean and one noisy, show that the estimated $T_{60}$ and equalization are both similar} \zhy{This may be difficult to do for both $T_{60}$ and EQ. We already trained with noise so this can be optional.}

% \subsection{Acoustic Matching}
% \zhy{Plan for this subsection: show how good our optimizer matches with target in Figure~\ref{fig:optimization_comparison}; show the optimized $T_{60}$ consistently matches targets in the sweep test shown in Figure~\ref{fig:sweep};}

%%%%%%%%%%%%%%%%%%%%%%%%%%%%%%%%%%%%%%%%%%%%%%%%%%%%%%%%%%%%%%%%%%%%%%%%%%%%%%%%
%%%%%%%%%%%%%%%%%%%%%%%%%%%%%%%%%%%%%%%%%%%%%%%%%%%%%%%%%%%%%%%%%%%%%%%%%%%%%%%%
\subsection{Comparisons}
We compare our work with two related projects, Schissler et al.~\cite{schissler2017acoustic} and Kim et al.~\cite{kim2019immersive},  where the high-level goal is similar to ours but the specific  approach is different.

\begin{figure}[tb]
    \centering
    \includegraphics[width=\linewidth]{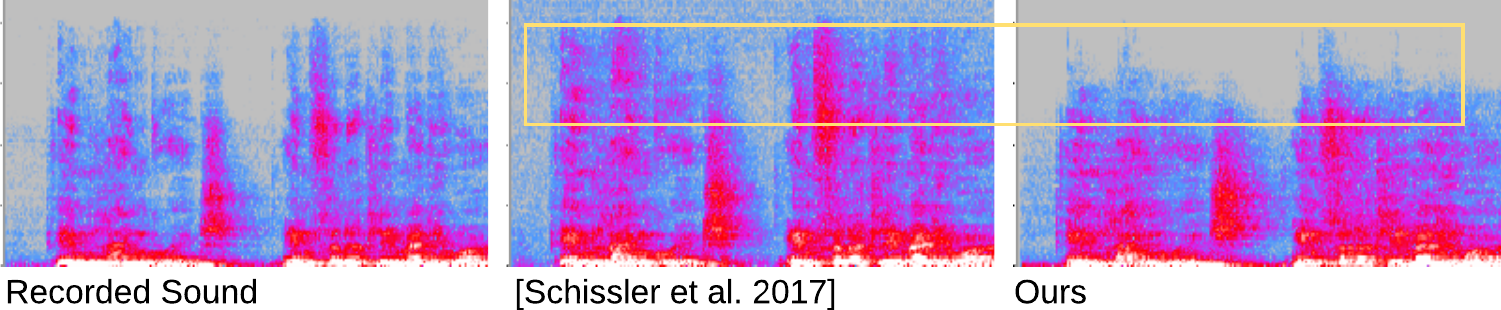}
    \caption{We show the effect of our equalization filtering on audio spectrograms, compared with Schissler et al.~\cite{schissler2017acoustic}.  In the highlighted region, we are able to better reproduce the fast decay in the high-frequency range, closely matching the recorded sound.}
    \label{fig:carl_comp}
\end{figure}

Material optimization is a key step in our method and Schissler et al.~\cite{schissler2017acoustic}.
One major difference is that we additionally compensate for wave effects explicitly with an equalization filter.
Figure~\ref{fig:carl_comp} shows the difference in spectrograms, where the high frequency equalization was not properly accounted for.
% \zhy{Carl's method requires multiple sets of measured IRs (3 in his paper) for joint optimization. It works for matching $T_{60}$. The main thing is they do not compensate wave effects. The difference in spectrogram can be seen in highlighted regions shown in Figure~\ref{fig:carl_comp}.}
Our method better replicates the rapid decay in the high frequency range. 
For audio comparison, please refer to our supplemental video.

\begin{figure}[tb]
    \centering
    \includegraphics[width=0.9\linewidth]{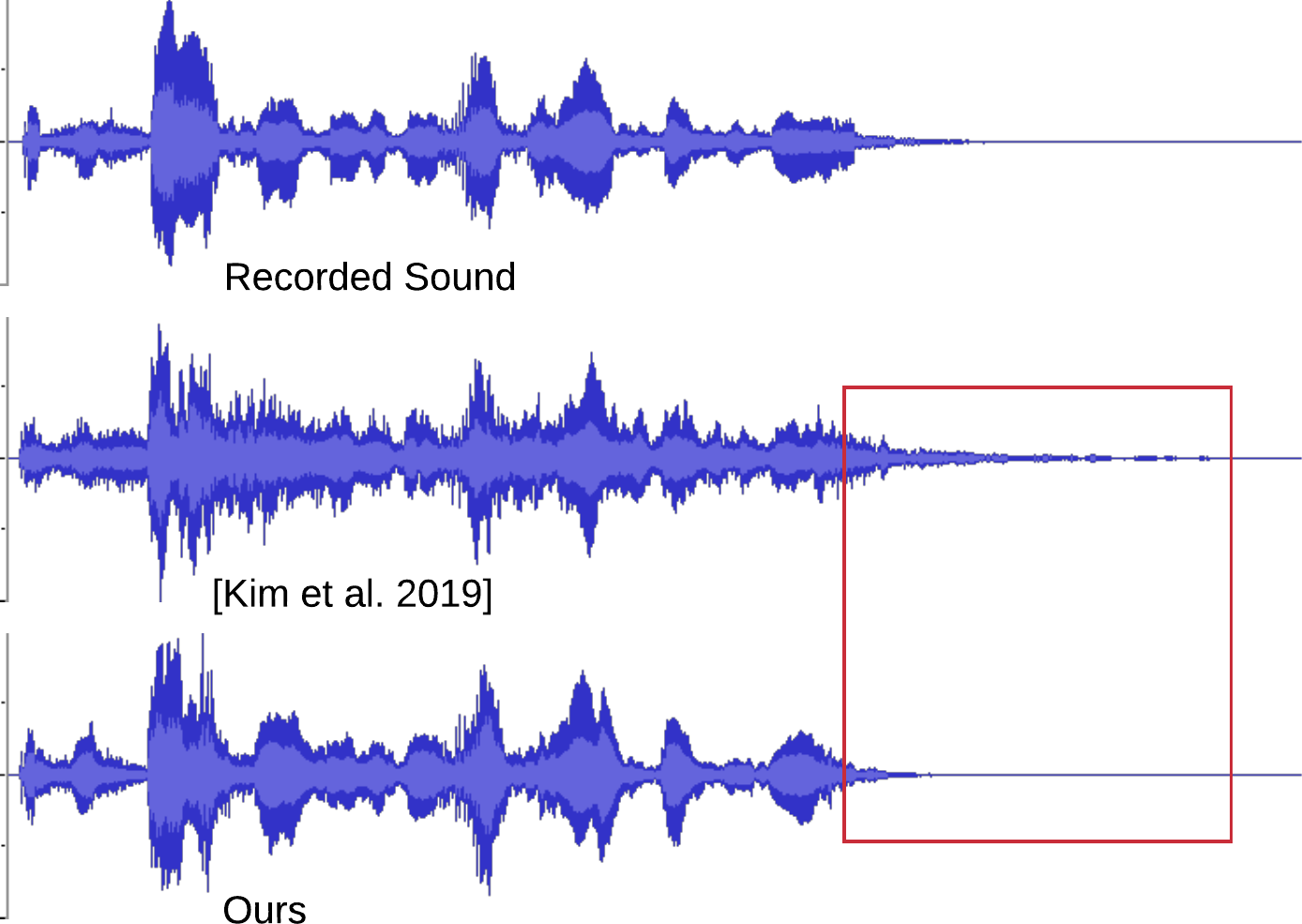}
    \caption{We demonstrate the importance on $T_{60}$ optimization on the audio amplitude waveform. 
    Our method optimizes the material parameters based on input audio and matches the tail shape and decay amplitude with the recorded sound, whereas the visual-based object materials from Kim et al.~\cite{kim2019immersive} failed to compensate for the audio effects.
    }
    \label{fig:kim_comp}
\end{figure}

We also want to highlight the importance of optimizing $T_{60}$.
In \cite{kim2019immersive}, a CNN is used for object-based material classification.
Default materials are assigned to a limited set of objects.
Without optimizing specifically for the audio objective, the resulting sound might not blend in seamlessly with the existing audio.
In Figure~\ref{fig:kim_comp}, we show that our method produces audio that matches the decay tail better,
whereas \cite{kim2019immersive} produces a longer reverb tail than the recorded ground truth. 

% \zhy{Kim et al's method does not require measured IR. They use two assembled 360 cameras to estimate depth map of the room, and run CNN for object based material classification. Default materials are assigned to a limited set of objects. Their goal should be plausible sound, but even that aspect is not well tested. Their results are somewhat close to our mid-anchor with wrong materials.}
% \todo{What's the comparison here? Did we optimize the materials and rerun the sim?}
% \zhy{Comparison would be Kim's method that relies on default material parameters vs our optimized materials, shown in Figure~\ref{fig:kim_comp} that their method results in longer reverb tail than groundtruth.}

% \zhy{We've said quite a lot about Li et al. in prior sections I wonder whether we want to repeat those here?} \todo{in that case, we can skip Li}

%%%%%%%%%%%%%%%%%%%%%%%%%%%%%%%%%%%%%%%%%%%%%%%%%%%%%%%%%%%%%%%%%%%%%%%%%%%%%%%%
%%%%%%%%%%%%%%%%%%%%%%%%%%%%%%%%%%%%%%%%%%%%%%%%%%%%%%%%%%%%%%%%%%%%%%%%%%%%%%%%
\subsection{Applications}
\label{sec:app}
% \begin{itemize}
%     \item single-image examples: components, $T_{60}$ only: 1) one room in Ding's paper, find picture and use room dimension from Ding's paper
%     2) Carl's small room with reconstruction
%     3) Kim et al comparison
%     \item video examples: 
%     1) Talking with virtual avatar (iphone / authenticity)  \url{https://www.youtube.com/watch?v=96Ivji-ph4A}
%     2) shadowing effects (whiteboard occlusion)
% \end{itemize}

% \todo{Figure~\ref{fig:comparisons}}: 
% \zhy{Compare our method with Kim et al. and Carl's work here. Cover video demo materials.}
\paragraph{Acoustic Matching in Videos}
Given a recorded video in an acoustic environment, our method can analyze the room acoustic properties from noisy, reverberant recorded audio in the video. 
The room geometry can be estimated from video~\cite{Bloesch_2018_CVPR}, if the user has no access to the room for measurement. 
During post-processing, we can simulate sound that is similar to the recorded sound in the room. 
Moreover, virtual characters or speakers, such as the ones shown in Figure~\ref{fig:teaser}, can be added to the video, generating sound that is consistent with the real-world environment.
%, enabling immersive remote virtual conferences.

\paragraph{Real-time Immersive Augmented Reality Audio}
Our method works in a real-time manner and can be integrated into modern AR systems. 
AR devices are capable of capturing real-world geometry, and can stream audio input to our pipeline. 
At interactive rates, we can optimize and update the material properties, and update the room EQ filter as well. 
Our method is not hardware-dependent and can be used with any AR device (which provides geometry and audio) to enable a more immersive listening experience.

\paragraph{Real-world Computer-Aided Acoustic Design}
Computer-aided design (CAD) software has been used for designing architecture acoustics, usually before construction is done, in a predictive manner~\cite{pelzer2014integrating,kleiner1990auralization}. 
But when given an existing real-world environment, it becomes challenging for traditional CAD software to adapt to current settings because acoustic measurement can be tedious and error-prone. 
By using our method, room materials and EQ properties can be estimated from simple input, and can be further fed to other acoustic design applications in order to improve the room acoustics such as material replacement, source and listener placement~\cite{morales2019receiver}, and soundproofing setup. 
% \zhy{will cite some CAD papers and acoustic design papers here}
% cited above in the first sentence

%%%%%%%%%%%%%%%%%%%%%%%%%%%%%%%%%%%%%%%%%%%%%%%%%%%%%%%%%%%%%%%%%%%%%%%%%%%%%%%%
%%%%%%%%%%%%%%%%%%%%%%%%%%%%%%%%%%%%%%%%%%%%%%%%%%%%%%%%%%%%%%%%%%%%%%%%%%%%%%%%

\section{Perceptual Evaluation}
\label{sec:study}
We perceptually evaluated our approach using a critical listening test. 
For this test, we studied the perceptual similarity of a reference speech recording with speech recordings convolved with simulated impulse responses. 
We used the same speech content for the reference and all stimuli under testing and evaluated how well we can reconstruct the same identical speech content in a given acoustic scene. 
This is useful for understanding the absolute performance of our approach compared to the ground truth results.

%%%%%%%%%%%%%%%%%%%%%%%%%%%%%%%%%%%%%%%%%%%%%%%%%%%%%%%%%%%%%%%%%%%%%%%%%%%%%%%%
%%%%%%%%%%%%%%%%%%%%%%%%%%%%%%%%%%%%%%%%%%%%%%%%%%%%%%%%%%%%%%%%%%%%%%%%%%%%%%%%
\subsection{Design and Procedure}
For our test, we adopted the multiple stimulus with hidden reference and anchor (MUSHRA) methodology from the ITU-R BS.1534-3 recommendation~\cite{series2014recommendation}. 
MUSHRA provides a protocol for the subjective assessment of intermediate quality level of audio systems~\cite{series2014recommendation} and has been adopted for a wide variety of audio processing tasks such as audio coding, source separation, and speech synthesis evaluation~\cite{schoeffler2015towards, cartwright2016fast}. 

\begin{figure}[tb]
\centering 
\includegraphics[width=0.9\linewidth]{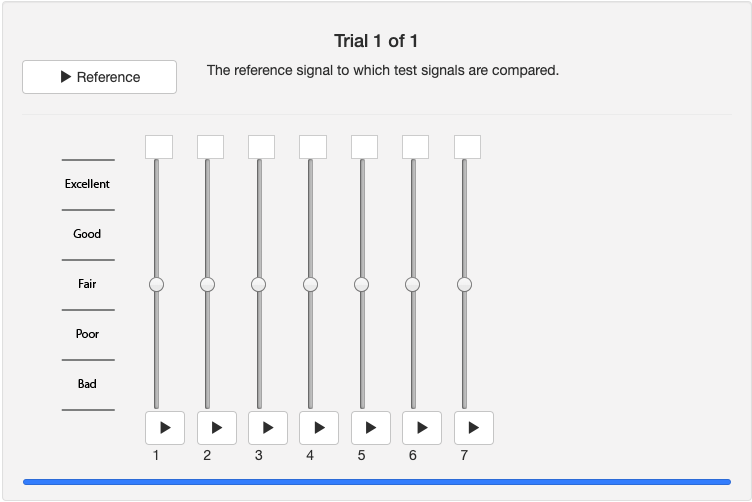}
\caption{A screenshot of MUSHRA-like web interface used in our user study. The design is from Cartwright et al.~\cite{cartwright2016fast}.}
\label{fig:mushralike}
\end{figure}

In a single MUSHRA trial, participants are presented with a high-quality reference signal and asked to compare the quality (or similarity) of three to twelve stimuli on a 0-100 point scale using a set of vertical sliders as shown in Figure~\ref{fig:mushralike}. 
The stimuli must contain a hidden reference (identical to the explicit reference), two anchor conditions -- low-quality and high-quality, and any additional conditions under study (maximum of nine). 
The hidden reference and anchors are used to help the participants calibrate their ratings relative to one another, as well as to filter out inaccurate assessors in a post-screening process. 
MUSHRA tests serve a similar purpose to mean opinion (MOS) score tests~\cite{mos2016method}, but requires fewer participants to obtain results that are statistically significant.

%To more quickly and easily perform our MUSHRA tests, w
We performed our studies using Amazon Mechanical Turk (AMT), resulting in a MUSHRA-like protocol~\cite{cartwright2016fast}. 
In recent years, web-based MUSHRA-like tests have become a standard methodology and have been shown to perform equivalently to full, in-person tests\cite{schoeffler2015towards, cartwright2016fast}.

%%%%%%%%%%%%%%%%%%%%%%%%%%%%%%%%%%%%%%%%%%%%%%%%%%%%%%%%%%%%%%%%%%%%%%%%%%%%%%%%
%%%%%%%%%%%%%%%%%%%%%%%%%%%%%%%%%%%%%%%%%%%%%%%%%%%%%%%%%%%%%%%%%%%%%%%%%%%%%%%%
\subsection{Participants}
We recruited 269 participants on AMT to rate one or more of our five acoustic scenes under testing following the approach proposed by Cartwright et al.~\cite{cartwright2016fast}. To increase the quality of the evaluation, we pre-screened the participants for our tests. 
To do this, we first required that all participants have a minimum number of 1000 approved Human Intelligence Task (HITs)  assignments and have had at least 97 percent of all assignments approved. 
Second, all participants must pass a hearing screening test to verify they are listening over devices with an adequate frequency response. 
This was performed by asking participants to listen to two separate eight second recordings consisting of a 55Hz tone, a 10kHz tone and zero to six tones of random frequency. 
If any user failed to count the number of tones correctly after two or more attempts, they were not allowed to proceed. Out of the 269 participants who attempted our test, 261 participants passed.

%%%%%%%%%%%%%%%%%%%%%%%%%%%%%%%%%%%%%%%%%%%%%%%%%%%%%%%%%%%%%%%%%%%%%%%%%%%%%%%%
%%%%%%%%%%%%%%%%%%%%%%%%%%%%%%%%%%%%%%%%%%%%%%%%%%%%%%%%%%%%%%%%%%%%%%%%%%%%%%%%
\subsection{Training}
After having passed our hearing screening test, each user was presented with a one page training test.
For this, the participant was provided two sets of recordings. 
The first set of training recordings consisted of three recordings: a reference, a low-quality anchor, and a high-quality anchor. 
The second set of training recordings consisted of the full set of recordings used for the given MUSHRA trail, albeit without the vertical sliders present. 
To proceed to the actual test, participants were required to listen to each recording in full.  
In total, we estimated the training time to be approximately two minutes.

%%%%%%%%%%%%%%%%%%%%%%%%%%%%%%%%%%%%%%%%%%%%%%%%%%%%%%%%%%%%%%%%%%%%%%%%%%%%%%%%
%%%%%%%%%%%%%%%%%%%%%%%%%%%%%%%%%%%%%%%%%%%%%%%%%%%%%%%%%%%%%%%%%%%%%%%%%%%%%%%%
\subsection{Stimuli}
For our test conditions, we simulated five different acoustic scenes. For each scene, a separate MUSHRA trial was created. In AMT language, each scene was presented as a separate HIT per user. For each MUSHRA trial or HIT, we tested the following stimuli: hidden reference, low-quality anchor, mid-quality anchor, baseline $T_{60}$, Baseline $T_{60}$+EQ, proposed $T_{60}$, and proposed $T_{60}$+EQ. 

As noted by the ITU-R BS.1534-3 specification~\cite{series2014recommendation}, both the reference and anchors have a significant effect on the test results, must resemble the artifacts from the systems, and must be designed carefully.  
For our work, we set the hidden reference as an identical copy of the explicit reference (required), which consisted of speech convolved with the ground truth IR for each acoustic scene. Then, we set the low-quality anchor to be completely anechoic, non-reverberated speech. We set the mid-quality anchor to be speech convolved with an impulse response with a 0.5 second $T_{60}$ (typical conference room) across frequencies, and uniform equalization. 

For our baseline comparison, we included two baseline approaches following previous work~\cite{Li:2018:360audio}. More specifically, our Baseline $T_{60}$ leverages the geometric acoustics method proposed by Cao et al.~\cite{cao2017bidirectional} as well as the materials analysis calibration method of Li et al.~\cite{Li:2018:360audio}. 
Our Baseline $T_{60}$+EQ extends this and includes the additional frequency equalization analysis~\cite{Li:2018:360audio}. 
These two baselines directly correspond to the proposed materials optimization (Proposed $T_{60}$) and equalization prediction subsystems (Proposed $T_{60}$+EQ) in our work.
The key difference is that we estimate the parameters necessary for both steps \emph{blindly from speech}.

%%%%%%%%%%%%%%%%%%%%%%%%%%%%%%%%%%%%%%%%%%%%%%%%%%%%%%%%%%%%%%%%%%%%%%%%%%%%%%%%
%%%%%%%%%%%%%%%%%%%%%%%%%%%%%%%%%%%%%%%%%%%%%%%%%%%%%%%%%%%%%%%%%%%%%%%%%%%%%%%%
%\subsection{Post-Filtering}
\subsection{User Study Results}

When we analyzed the results of our listening test, we post-filtered the results following the ITU-R BS.1534-3 specification~\cite{series2014recommendation}. 
%, Section 4.1.2
More specifically, we excluded assessors if they
\begin{itemize}
  \setlength\itemsep{0em}
    \item rated the hidden reference condition for $> 15\%$ of the test items lower than a score of 90
    \item or, rated the mid-range (or low-range) anchor for more than $15\%$ of the test items higher than a score of 90.
\end{itemize}
Using this post-filtering, we reduce our collected data down to 70 unique participants and 108 unique test trials, spread across our five acoustic scene conditions. Among these participants, 24 are females and 46 are males, with an average age of 36.0 and a standard deviation of 10.2 years.

%%%%%%%%%%%%%%%%%%%%%%%%%%%%%%%%%%%%%%%%%%%%%%%%%%%%%%%%%%%%%%%%%%%%%%%%%%%%%%%%
\begin{figure}[tb]
\centering 
\includegraphics[trim={0.5cm 0 0 0},clip,width=\linewidth]{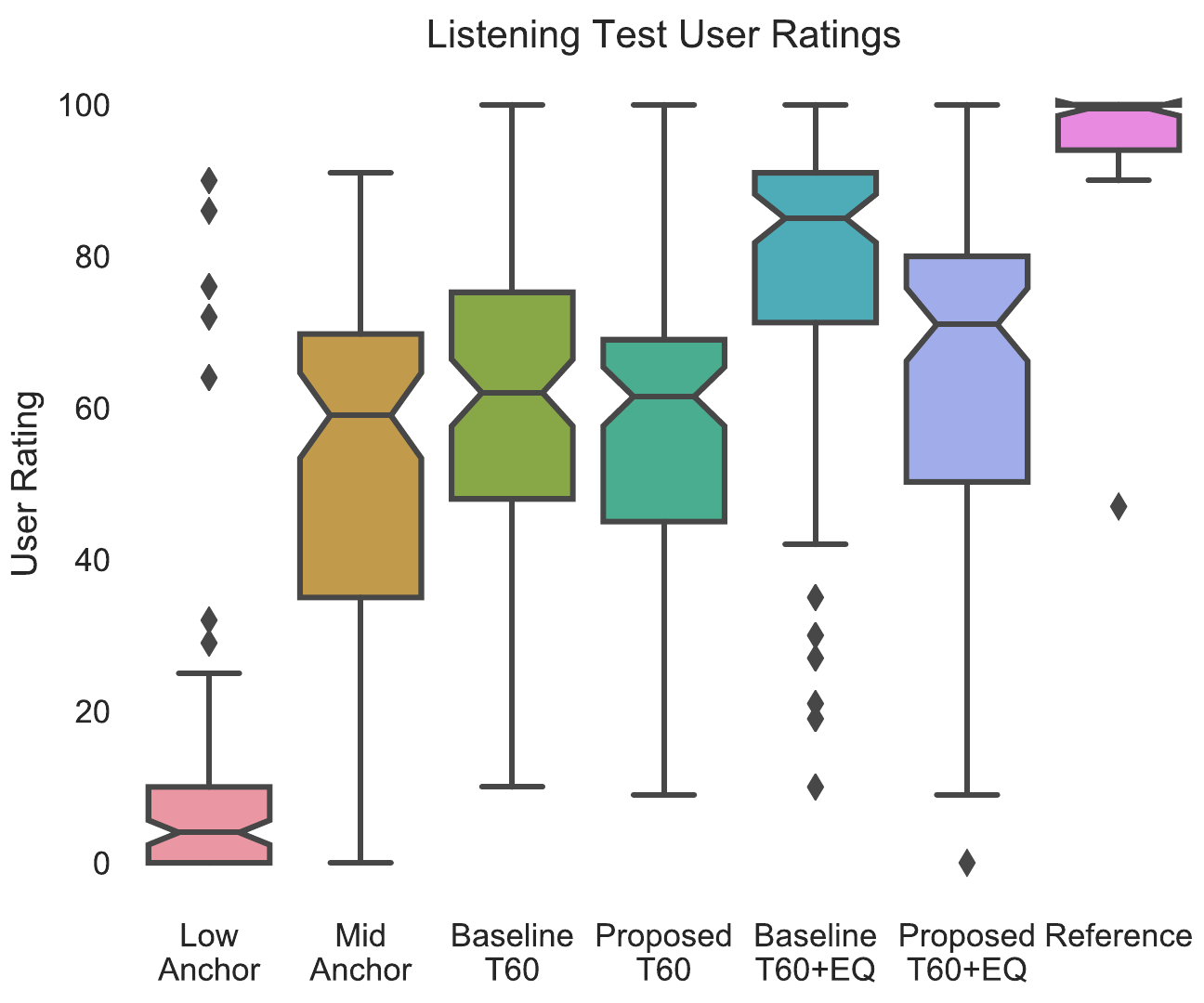}
\caption{Box plot results for our listening test. Participants were asked to rate how similar each recording was to the explicit reference. 
All recordings have the same content, but different acoustic conditions.
Note our proposed $T_{60}$ and $T_{60}$+EQ are both better than the Mid-Anchor by a statistically significant amount ($\approx$10 rating points on a 100 point scale).
}
\label{fig:authenticity}
\end{figure}
We show the box plots of our results in Figure~\ref{fig:authenticity}. The median ratings for each stimulus include: Baseline $T_{60}$ (62.0), Baseline $T_{60}$+EQ (85.0), Low-Anchor (40.5), Mid-Anchor (59.0), Proposed $T_{60}$ (61.5),  Proposed $T_{60}$+EQ (71.0), and Hidden Reference (99.5).
As seen, the Low-Anchor and Hidden Reference outline the range of user scores for our test.
In terms of baseline approaches, the Proposed $T_{60}$+EQ method achieves the highest overall listening test performance.
We then see that our proposed $T_{60}$ method and $T_{60}$+EQ method outperform the mid-anchor. Our proposed $T_{60}$ method is comparable to the baseline $T_{60}$ method, and our proposed $T_{60}$+EQ method outperforms our proposed $T_{60}$-only method. %We also note, our proposed $T_{60}$+EQ method does not outperform the baseline $T_{60}$+EQ. This is expected, however, because the baseline approach has an exact estimate by using an explicit, measured IR. Our method does not have this advantage.

To understand the statistical significance, we performed a repeated measures analysis of variance (ANOVA) to compare the effect of our stimuli on user ratings. The Hidden Reference and Low-Anchor are for calibration and filtering purposes and are not included in the following statistical tests, leaving 5 groups for comparison. Bartlett's test did not show a violation of homogeneity of variances ($\chi^2 = 4.68, p = 0.32$). A one-way repeated measures ANOVA shows significant differences ($F(4,372)=29.24, p<0.01$) among group mean ratings. To identify the source of differences, we further conduct multiple post-hoc paired t-tests with Bonferroni correction~\cite{holm1979simple}. We are able to observe following results: a) There is no significant difference between Baseline $T_{60}$ and Proposed $T_{60}$ ($t(186)=-1.72, p=0.35$), suggesting that we cannot reject the null hypothesis of identical average scores between prior work (which uses manually measured IRs) and our work; b) There is a significant difference between Baseline $T_{60}$+EQ and Proposed $T_{60}$+EQ ($t(186)=-5.09, p<0.01$), suggesting our EQ method has a statistically different average (lower); c) There is a significant difference between Proposed $T_{60}$ and Proposed $T_{60}$+EQ ($t(186)=-2.91, p=0.02$), suggesting our EQ method significantly improves performance compared to our proposed $T_{60}$-only subsystem; d) There is a significant difference between Mid-Anchor and Proposed $T_{60}$+EQ ($t(186)=-3.78, p<0.01$), suggesting our method is statistically different (higher performing) on average than simply using an average room $T_{60}$ and uniform equalization.

% To understand the statistical significance, we perform paired t-tests between stimuli pairs. The p-value between Baseline $T_{60}$ and Proposed $T_{60}$ is 0.09, suggesting that we cannot reject the null hypothesis of identical average scores between prior work (which uses manually measured IRs) and our work. The p-value of Baseline $T_{60}$+EQ and Proposed $T_{60}$+EQ, however, is 1.85e-6, suggesting our EQ method has a statistically different average (lower). The p-value of Proposed $T_{60}$ and Proposed $T_{60}$+EQ, however, is 0.004, suggesting our EQ method does significantly improve performance compared to our proposed $T_{60}$-only subsystem. We also note that the p-value of the Mid-Anchor and Proposed $T_{60}$+EQ is 0.0002, suggesting our method is statistically different (higher performing) on average than simply using an average room $T_{60}$ and uniform equalization. 

In summary, we see that our proposed $T_{60}$ computation method is comparable to prior work, albeit we perform such estimation directly from a short speech recording rather than relying on intrusive IR measurement schemes. Further, our proposed complete system (Proposed $T_{60}$+EQ) outperforms both the mid-anchor and proposed $T_{60}$ system alone, demonstrating the value of EQ estimation. Finally, we note our proposed $T_{60}$+EQ method does not perform as well as prior work, largely due to the EQ estimation subsystem. This result, however, is expected as prior work requires manual IR measurements, which result in perfect EQ estimation. This is in contrast to our work, which directly estimates both $T_{60}$ and EQ parameters from recorded speech, enabling a drastically improved interaction paradigm for matching acoustics in several applications.   

\section{Conclusion and Future Work}

We present a new pipeline %\zhy{is this claim too strong (unsafe)?}  changed from `first` to `a`
to estimate, optimize, and render immersive audio in video and mixed reality applications. We present novel algorithms to estimate two important  acoustic environment characteristics -- the frequency-dependent reverberation time and equalization filter of a room.
%By removing the requirement on separate impulse response recording, we significantly expand the application scenarios to be backward compatible with all existing videos.
Our multi-band octave-based prediction model works in tandem with our equalization augmentation and provides robust input to our improved materials optimization algorithm.
Our user study validates the perceptual importance of our method.
To the best of our knowledge, our method is the first method to predict IR equalization from raw speech data and validate its accuracy.
%and compares against state-of-the-art baseline methods.
%While not quite as high quality as baselines, our method works \emph{blindly} from existing speech recordings, increasing the ease of use and possible application domains.
% \todo{mention user study?}
% In summary, our system makes it possible to author audio in much wider spectrum of visual contents. 

\paragraph{Limitations and Future Work.} 
To achieve a perfect acoustic match, one would expect the real-world validation error to be zero. 
In reality, zero error is only a sufficient but not necessary condition.
In our evaluation tests, we observe that small validation errors still allow for plausible acoustic matching.
While reducing the prediction error is an important direction, it is also useful to investigate the perceptual error threshold for acoustic matching for different tasks or applications.
Moreover, temporal prediction coherence is not in our evaluation process. This implies that given a sliding windows of audio recordings, our model might predict temporally incoherent $T_{60}$ values. One interesting problem is to utilize this coherence to improve the prediction accuracy as a future direction.

Modeling real-world characteristics in simulation is a non-trivial task -- 
as in previous work along this line, our simulator does not fully recreate the real world in terms of precise details.
For example, we did not consider the speaker or microphone response curve in our simulation.
In addition, sound sources are modeled as omnidirectional sources~\cite{cao2017bidirectional}, where real sources exhibit certain directional patterns.
It remains an open research challenge to perfectly replicate and simulate our real world in a simulator.

Like all data-driven methods, our learned model performs best on the same kind of data on which it was trained.
Augmentation is useful because it generalizes the existing dataset so that the learned model can extrapolate to unseen data.
However, defining the range of augmentation is not straightforward. 
We set the MIT IR dataset as the baseline for our augmentation process.
In certain cases, this assumption might not generalize well to estimate the extreme room acoustics. We need to design better and more universal augmentation training algorithms.
%to address this challenge.
Our method focused on estimation from speech signals, due to their pervasiveness and importance. 
It would be useful to explore how well the estimation could work on other audio domains, especially when interested in frequency ranges outside typical human speech. 
This could further increase the usefulness of our method, e.g., if we could estimate acoustic properties from ambient/HVAC noise instead of requiring a speech signal.

\acknowledgments{
The authors would like to thank Chunxiao Cao for sharing the bidirectional sound simulation code, James Traer for sharing the MIT IR dataset, and anonymous reviewers for their constructive feedback. This work was supported in part by ARO grant W911NF-18-1-0313, NSF grant \#1910940, Adobe Research, Facebook, and Intel.}

\bibliographystyle{abbrv-doi}

\balance
\bibliography{AR-audio}
\end{document}